\documentclass[fleqn,usenatbib]{mnras}

\usepackage{newtxtext,newtxmath}

\usepackage[T1]{fontenc}
\usepackage{ae,aecompl}

\usepackage{graphicx}	
\usepackage{amsmath}	
\usepackage{pdflscape}	
\usepackage{threeparttable}
\usepackage{ulem} 

\title[QP Kernel and Physical Parameters]{Quasi-periodic Gaussian Processes for stellar activity: from physical to kernel parameters
}

\author[Nicholson \& Aigrain]{
B. A. Nicholson,$^{1,2}$\thanks{E-mail: belinda.nicholson@physics.ox.ac.uk}
and S. Aigrain,$^{1}$
\\
$^{1}$Sub-department of Astrophysics, University of Oxford, Keble Rd, Oxford, United Kingdom, OX13RH\\
$^{2}$University of Southern Queensland, Centre for Astrophysics, Toowoomba, Australia, 4350\\
}

\date{Accepted 25 July 2022}

\pubyear{2022}

\begin{document}
\label{firstpage}
\pagerange{\pageref{firstpage}--\pageref{lastpage}}
\maketitle

\begin{abstract}
In recent years, Gaussian Process (GP) regression has become widely used to analyse stellar and exoplanet time-series data sets. For spotted stars, the most popular GP covariance function is the quasi-periodic (QP) kernel, whose the hyperparameters of the GP have a plausible interpretation in terms of physical properties of the star and spots. In this paper, we test the reliability of this interpretation by modelling data simulated using a spot model using a QP GP, and the recently proposed quasi-periodic plus cosine (QPC) GP, comparing the posterior distributions of the GP hyperparameters to the input parameters of the spot model. We find excellent agreement between the input stellar rotation period and the QP and QPC GP period, and very good agreement between the spot decay timescale and the length scale of the squared exponential term. We also compare the hyperparameters derived from light and radial velocity (RV) curves for a given star, finding that the period and evolution timescales are in good agreement. However, the harmonic complexity of the GP, while displaying no clear correlation with the spot properties in our simulations, is systematically higher for the RV than for the light curve data. Finally, for the QP kernel, we investigate the impact of noise and time-sampling on the hyperparameters in the case of RVs. Our results indicate that good coverage of rotation period and spot evolution time-scales is more important than the total number of points, and noise characteristics govern the harmonic complexity. 
\end{abstract}

\begin{keywords}
stars: activity -- methods: data analysis -- techniques: photometric -- techniques: radial velocities
\end{keywords}



\section{Introduction}

Gaussian Process (GP) regression has become an increasingly popular method for analysing stellar activity signals in photometric and radial velocity (RV) time-series data. In particular, GP models based on quasi-periodic kernel functions
are able to reproduce the rotationally-modulated signatures of evolving magnetically active regions remarkably well \citep[see e.g.][]{Aigrain2012,Haywood2014,Angus2018}. Quasi-periodic GP models have been applied extensively to stellar time-series data, both to mitigate the impact of stellar activity on planet detection or characterisation \citep[see e.g.][and numerous later references]{Haywood2014,2015MNRAS.452.2269R,Grunblatt2015,2015ApJ...800...46B}, and to study stellar activity signals \textit{per se} \citep[see e.g.][]{Angus2018,2021RNAAS...5...51B}.

There are many ways to construct a quasi-periodic kernel function for a GP model, but one such kernel has become especially popular for modelling stellar activity signals, to the extent that, in the exoplanet community, it is often referred to simply as the Quasi-Periodic (QP) kernel. The QP kernel is constructed by multiplying a periodic term, consisting of the exponential of a sine-squared function, with a decaying envelope, consisting of a squared exponential function. With only 4 parameters, this kernel function can produce signals with different amplitudes, periods, evolution time and degree of harmonic complexity. Random samples drawn from a QP GP prior bear a striking resemblance to observed stellar light curves \citep[see e.g.][]{2015MNRAS.452.2269R}.

A QP kernel isn't always the best choice of kernel to model stellar activity signals. For example, \citet{Gilbertson2020} show that simpler, a-periodic kernels can in some circumstances perform somewhat better in terms of activity mitigation in RV time-series. However, the QP kernel explicitly encodes the commonly held belief that the underlying signal should be quasi-periodic, even if its periodic nature is not necessarily apparent in noisy and/or sparsely sampled data. Furthermore, most of the parameters of the QP kernel (known as the hyper-parameters, or HPs, of the GP) admit a fairly natural interpretation in terms of the physical properties of the active regions, such as their rotation rates and lifetimes. The primary goal of this study is to investigate, using simulated data, the extent to which this "natural" interpretation is robust and can be used to guide modelling decisions.

Observed light and RV curves of active stars often contain significant power not only at the stellar rotation period, but at its first few harmonics \citep[see e.g.][]{Aigrain2012}, a feature which is not naturally reproduced by the standard QP kernel. Consequently, in recent years a number of other quasi-periodic kernels have been investigated, which explicitly generate power at half the rotation period, including the "QP plus cosine" (hereafter QPC) kernel introduced by \citet[][hereafter P21]{Perger2021}, given by Equation~\ref{eq:QPC}, and the "rotation term" implemented in the {\sc celerite} package \citep{celerite}. After experimenting with both of these, we decided to include the former in the present study. Achieving convergence with, and robustly interpreting the parameters of, GP models based on the {\sc celerite} rotation term is somewhat more challenging with our model data, and we defer that to a separate paper.

While GP models using the QP and QPC kernels provide a useful description of common observables, they are still very much phenomenological. A more physically motivated family of GP models for active star light curves was recently introduced by \citet{Luger2021b}. Their {\sc starry-process} framework places a GP prior on the distribution of active regions on the stellar surface, which is described using a spherical harmonic decomposition. However, this relatively new approach has not yet been widely used, nor has it been extended to RV data, placing it outside the scope of the present work.

The fact that a QP GP can be used to recover stellar rotation periods from light curves has already been demonstrated \citep[see e.g.][]{Angus2018}. However, the behaviour of the other HPs, in particular the parameter controlling the evolution of the signal, has not yet been investigated in detail. A number of recent studies have shown that it is possible to measure active region lifetimes from photometry, for example by fitting a function to the decaying envelope auto-correlation function (ACF) of the light curve \citep{Giles2017,Santos2021,Basri2021}. These studies have shown that, provided the light curve duration significantly exceeds the active region lifetime, the ACF decay time is roughly proportional to the active region lifetime, but the constant of proportionality is not unity. We would expect the results obtained with a GP model, which effectively is a more direct way of modelling the covariance, to be consistent with those obtained by fitting the ACF, but this has yet to be demonstrated.

P21 analysed simulated RVs with four different spot distributions with GP models using QP, QPC and {\sc celerite} rotation kernels. They were able to recover the input spot evolution time, but their study considered only a single value of the rotation period and evolution time. In the present work, we extend this to a much broader parameter space, varying the rotation period, spot lifetime and number of spots.

A common application of the QP GP is in filtering stellar activity from RV time series. When following up planet candidates detected in transit, it is common practice to `train' the the GP on the light curve data first \citep[see e.g.][]{Haywood2014, theRADVELpaper, 2015MNRAS.452.2269R}. The posterior distribution over the HPs of the QP GP is then used as a prior for the RV analysis. However, this may not always be advisable, for two distinct reasons. First, the photometric data, if available, is rarely contemporaneous with the RVs, and can be taken months or even years apart. Using a 70-year compilation of solar irradiance data, \citet{Kosiarek2020} showed that all the HPs of a QP GP fit to the light-curve of a Sun-like star can vary significantly over the solar cycle. This implies that it may be dangerous to train a QP GP on light curve data before applying it to RV data, if the two were taken too far apart. Furthermore, their analysis of simultaneous SORCE photometry and HARPS-N RVs found consistent periods and evolution time-scales, but significant differences in the length-scale parameter that controls the complexity of the signal within a period (often referred to as the harmonic complexity). This effect was also seen in simulated data by \cite{PyanetiII}. It is related to the fact that the RV variations depend not only on the projected area of the active regions and their contrast, but also on their local line-of-sight velocity. This underpins the so-called FFprime framework for predicting RV variations from photometry \citep{Aigrain2012}, as well as latent variable models where RVs are modelled jointly with activity indicators extracted from the same spectra, as linear combinations of an underlying GP and its time-derivative(s) \citep[see e.g.][]{2015MNRAS.452.2269R,Jones2017,S+LEAF2}. However, as it is still quite common in the literature to model RVs alone using a QP GP trained on photometry, it is important to test the robustness of this approach.

In this work, we first use a simple spot model to simulate simulate tightly-sampled, noise-less photometric and RV time-series, and then model each independently with a QP or QPC GP. This allows us to test the extent to which the HPs of these models relate to the physical parameters of the simulations, and how closely the results based on photometry match each other. Doing this using highly idealised data enables us to explore the differences between these two types of observations at a fundamental level, removing the effects of noise and time-sampling. As a second step, we then degrade the time-sampling of the RV data, and add white noise, to make the simulations more realistic (since RV observations are typically ground-based). This allows us to investigate how time-sampling in particular affects the recovery of the GP HPs, in the case when space-based photometry is not available (i.e.\ for "pure RV" surveys, rather than for transit follow-up).

The remainder of this paper is organised as follows. In Section~\ref{sec:theGP}, we define the QP and QPC kernels used in this work, and fit simulated time-series generated from the GP prior itself. This allows us to test for degeneracies inherent to the GP models. In Section~\ref{sec:QPofLC}, we define the spot model used to simulate stellar light and RV curves, and examine the properties of the HPs obtained from the simulated light curves. The GP fits to the idealised RV datasets are discussed in Section~\ref{sec:GPofRV}, and the effect of degrading the time-sampling of the RVs are explored in Section~\ref{sec:RV_resamp}. Finally, we summarise and discuss the findings of this work in Section~\ref{sec:DiscConc}. 

\section{Quasi-periodic Gaussian Process models}
\label{sec:theGP}

GP regression is a powerful and flexible method for analysing data containing correlated noise. Rather than fitting a function to data, GP regression instead fits the correlation between data points. At its heart, a GP is defined by its kernel covariance function, $k$, which describes the extent to which two observations are correlated with each other, usually as a function of their distance in some input space (for time-series, the time interval separating them). For the purpose of modelling stellar activity signals in astronomical time-series, a quasi-periodic covariance function is commonly chosen. In the most general sense, a quasi-periodic kernel is usually obtained by multiplying a periodic term with a decaying envelope.

\subsection{The Quasi-Periodic kernel}

{In the standard QP kernel most commonly used for stellar activity modelling, the periodic term is given by the exponential of a sine-squared function, while the evolution is described by a squared exponential term. Formally, the QP kernel is defined as:}
\begin{equation}
  \label{eq:QP}
	k_{QP}(t,t') = A \exp\left[-\Gamma \sin^2\Big(\frac{\pi(t-t')}{P}\Big) - \frac{(t-t')^2}{2 l^2}\right], 
\end{equation}
where $A$ is the variance, $\Gamma$ is the scale factor (termed `harmonic complexity') and  $P$ the period of the sine-squared term, and $l$ is the `evolution  timescale' of the squared exponential term. These terms are referred to as the `hyperparameters' of the GP so as to differentiate this process from fitting a parametric function to the data. For further information of the quasi-periodic kernel GP and on GP's in general, we refer the reader to \cite{Rasmussen2006} {and} \cite{Roberts2013}. 

The QP GP used in this work is implemented using the {\sc George} python package \citep{Ambikasaran2015} by combining the {\sc ExpSine2Kernel} and {\sc ExpSquaredKernel} built-in kernel functions. The {\sc George} kernels use the hyperparameters values in natural log as input, with the exception of the $\Gamma$ term, and reparameterise the denominator of the squared exponential term as $m=2l^2$, called the `metric'. We also solve for the mean function of the data, in this case a constant, as this reflects the practice of fitting underlying functions (such as a Keplerian orbit in the case of RVs) to the data.

\subsection{The Quasi-Periodic plus Cosine kernel}

P21 introduced the idea of adding a cosine term to the periodic component of the QP kernel, with a period equal to half that of the sine-squared term, to better handle cases where the distribution of spots on the stellar surface gives rise to signal the first harmonic of the stellar rotation period (evidenced by a secondary peak in the auto-correlation function at half the rotation period). Mathematically, this "Quasi-Periodic plus Cosine" (QPC) kernel is defined as: 
\begin{multline}
  \label{eq:QPC}
    k_{QPC}(t,t') = A \exp\left[- \frac{(t-t')^2}{2 l^2}\right] \times \\ \left(\exp \left[-\Gamma \sin^2\left(\frac{\pi(t-t')}{P}\right)\right] + f\cos\left(\frac{4\pi(t-t')}{P}\right)\right) , 
\end{multline}
where $A$, $\Gamma$, $P$ and $l$ retain the same definitions as in the QP kernel, and we have introduced an additional hyperparameter $f$, which controls the amplitude of the cosine term relative to the sine-squared term. As $f$ tends to zero, the behaviour of $k_{QPC}$ tends to that of $k_{QP}$. 

In this work we define this kernel again in {\sc George} by with the addition of a {\sc CosineKernel} term to our {\sc George} QP kernel.

\subsection{Exploring inherent degeneracies in the hyperparameters}
\label{subsec:QPGPStability}

The first step in assessing the ability of a GP to recover physical parameters of a star is to investigate the stability of the GP fit itself and explore any inherent degeneracies in fitting a QP or QPC kernel. To test this, we generate a number of `light curves' by drawing samples from {our GP prior}, and fit these sample curves to see if we recover the input GP hyperparameters. {This step also serves to validate our fitting and posterior exploration procedure. The bulk of this analysis was carried out using the QP kernel, but we also discuss the QPC kernel and its extra hyperparameter towards the end of the section.}

Our sample curves were generated over a range of hyperparameters, with 20 sample curves per set of hyperparameters. Each sample curve spans 100 days, with a sampling of three, evenly spaced points per day, and a fixed amplitude of 1. The grid of hyperparameters contain period values of 2, 10, 30 and 60 days, $\Gamma$ values of 0.5 and 2 and $l$ values of a 1, 3 and 10 times the period. These values were chosen as they represent the range of values typically found in the analysis of stellar light curves across a range of activity levels. 

Markov Chain Monte Carlo (MCMC) was used to sample the posterior distribution of the hyperparameters for each sample curve, using the {\sc emcee} python package \citep{Foreman-Mackey2013}. The MCMC was set up with 80 walkers in a Gaussian distribution about the sample curve input values, with a broad uniform prior used for each hyperparameter across all sample curves (see Table \ref{tab:GP_Sim_priors}). There was an initial burn in of 100 steps, from which a tight Gaussian distribution around the highest probability state was used to re-initialise the walkers. The main MCMC was then run until the walkers had converged on a solution or a maximum of 8000 iterations was reached. The convergence of the MCMC walkers was checked every 100 steps, and a sample was deemed to have converged if the auto-correlation length of all chains was more than 60 times the number of elapsed iterations and had not changed by more than $1\%$ in the last 100 steps. For each of the MCMC solutions a `burn-in' of twice the maximum chain auto-correlation length was removed, and the chains were thinned by half the minimum chain auto-correlation length to ensure an unbiased sample.  The optimum solution for each hyperparameter was taken as the median of the posterior distribution, and the uncertainty as difference between the 50th and 16th percentile for the lower bound and the 84th and 50th percentile for the upper.  

\begin{table}
\centering
  \caption{Priors placed on the parameters and QP GP hyperparameters for the MCMC fit of the sample curves drawn from sets of QP GPs.}
  \label{tab:GP_Sim_priors}
  \begin{tabular}{ll}
  \hline
  \hline
	Input Parameter			& Prior$^{(\mathrm{a})}$\\
	\hline
	$\ln(A)$            & $\mathcal{U}[-30,30]$\\
	$\ln (P)$           & $\mathcal{U}[-15,15]$ \\
    $\Gamma$            & $\mathcal{U}[0,15]$\\
    $\ln (2l^2)$        & $\mathcal{U}[-30,30]$\\
    Mean                & $\mathcal{U}[-15,15]$\\
    \hline
  \end{tabular}
  \begin{tablenotes}\footnotesize
  \item \textit{Note} -- $^{(\rm{a})}$ $\mathcal{U}[a,b]$ refers to uniform priors between $a$ and $b$.
    \end{tablenotes}
\end{table}

\begin{figure*}
	\includegraphics[width=1.0\textwidth]{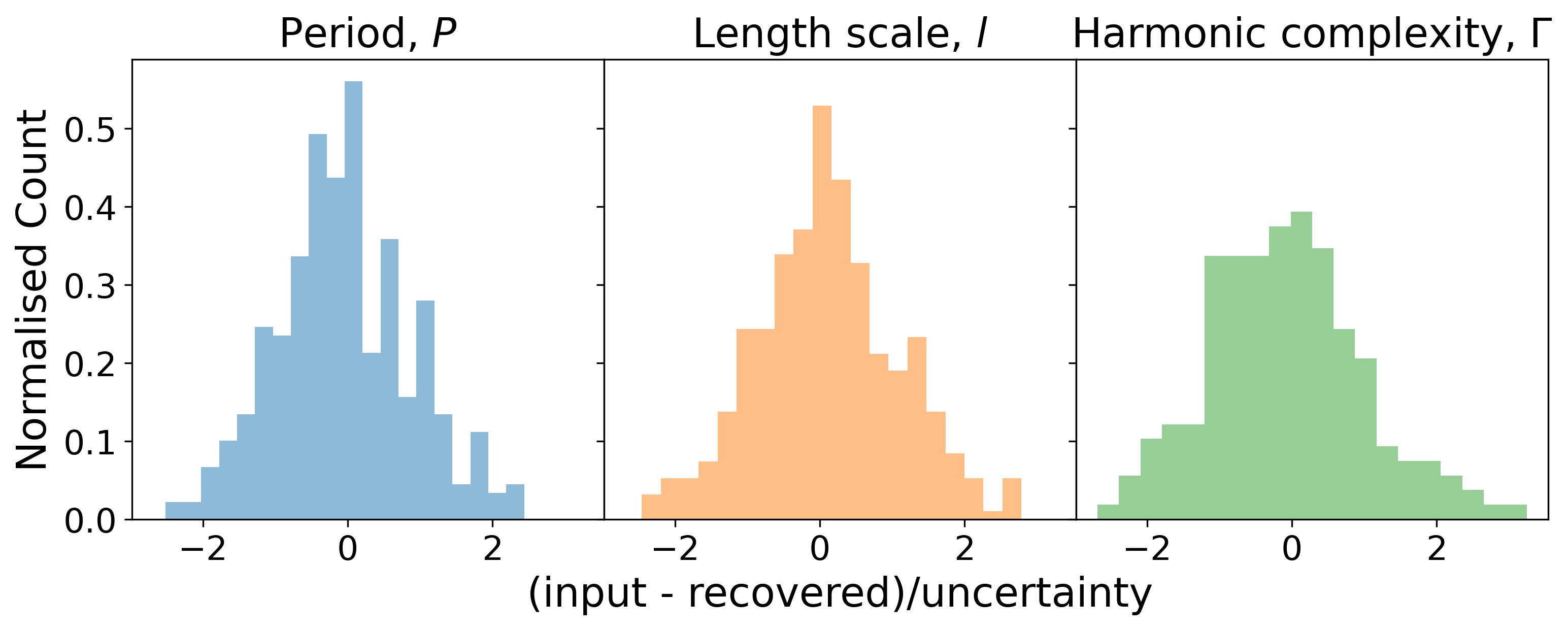}
    \caption{Recovery of input {QP }GP hyperparameters: each panel shows a histogram of the difference between the input and output parameter values divided by the uncertainty of the output solution for each of the varied hyperparameters: Period ($P$, left), length scale ($l$, middle), and harmonic complexity ($\Gamma$, right). }
    \label{fig:QP_Hists}
\end{figure*}

Figure \ref{fig:QP_Hists} shows how closely the input values of the period (left), $l$ (middle) and $\Gamma$ (right) were recovered by the GP, relative to the uncertainties calculated from the MCMC posterior distributions. Performing a Kolmogorov-Smirnov test on the distribution of (input-output)/uncertainty for each of period, $l$ and $\Gamma$, we find D statistic values of 0.081, 0.068 and 0.058, and p-values of 0.016, 0.071 and 0.18, respectively. The D statistic values are close to 0 (normal distribution), and the p-vales are not significant enough to reject the the null hypothesis that these distributions are drawn from a standard normal distribution. In other words, all three distributions are consistent with a Gaussian with a mean around zero and a standard deviation of 1. This means that, within the limits of statistical uncertainty, the QP GP is non-degenerate and stable: the parameters that go in are the parameters that will come out.

\begin{figure}
	\includegraphics[width=0.5\textwidth]{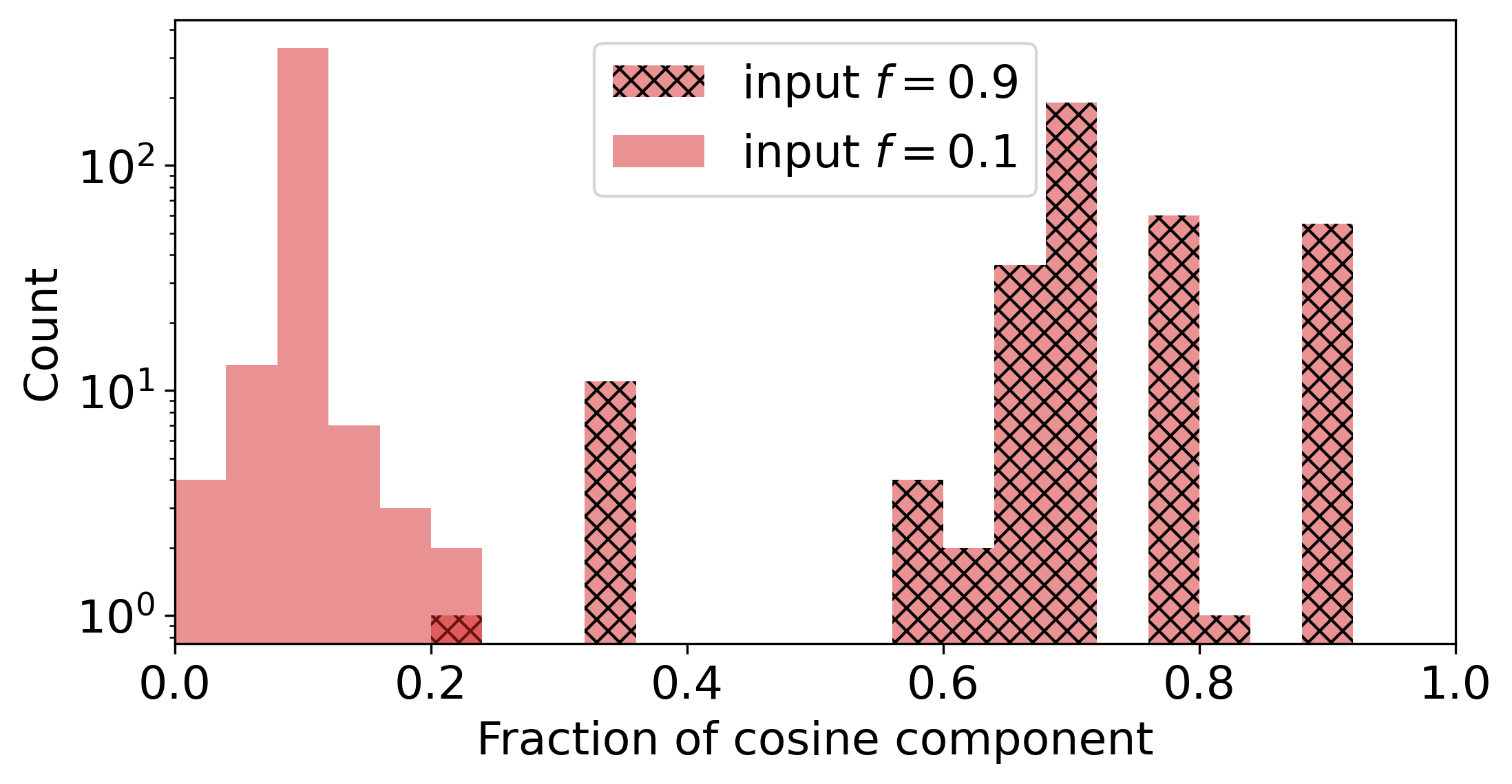}
    \caption{{Histograms of the recovered cosine term fraction $f$ from a QPC GP for to sample curves drawn from a QPC GP with a high $f$ value of 0.9 (hatched), and low $f$ value of 0.1 (solid).}} 
    \label{fig:QPC_hists}
\end{figure}

We then tested in the same manner a set of light curves generated from a QPC kernel. The input values for $P$, $\Gamma$ and $l$ were the same as in the QP case. We tested two values of $f$,  0.1 and 0.9, and adopted a uniform prior on $\log(f)$ between $\log(10^{-4})$ and 0 when fitting the samples. We find the same results for period, $l$ and $\Gamma$ as with the pure QP case: the input hyperparameters are all well-recovered. The fraction term, $f$, was only recovered in 7 cases within 3-sigma, all of which had high $f$. Figure \ref{fig:QPC_hists} shows histograms of the recovered $f$ values for the low (solid) and high (hatched) input values. For the low values of $f$, the QPC GP recover recovers the $f$ values, but not as well as it does for the other parameters. The high-$f$ value cases are not well recovered, despite these cases all having well recovered period, $l$ and $\Gamma$ values.  As seen in \ref{fig:QPC_hists}, the recovered $f$ for these high-$f$ cases range from 0.2 to 0.9, with a median value of 0.7. This indicates that the QPC GP is only weakly sensitive to the true fraction term on the cosine component.

\section{{Modelling spotted star light curves with quasi-periodic GPs}}
\label{sec:QPofLC}

In this section we describe how we simulate light curves for rotating stars with evolving active regions, and then fit them with quasi-periodic GPs using the kernels described in Section~\ref{sec:theGP}.

\subsection{Simulating spotted star light curves}

To test the correlation of the QP GP hyperparameters with physical characteristics of a star, we generate 100 simulated light curves using a simple spot model implemented in the \texttt{PySpot} package\footnote{See \url{https://github.com/saigrain/pyspot}.}\citep{aigrain_suzanne_2021_5654179}, which is a simplified version of the model used in the \textit{Kepler} rotation and differential rotation blind exercise presented in \cite{2015MNRAS.450.3211A}. We first define the parameters of the star and the spots in each simulation, then use these simulations to produce light and RV curves. 

All simulations last $T=250$ days, with 2 equally spaced observations per day, and all the stars are seen equator on, and have the same radius as the Sun. For each simulation, we draw the stellar rotation period, $P_{\rm rot}$, the average number of spots present on the stellar surface at any given time, $N_{\rm Spot}$, and the ratio of the characteristic life-time of the spots to the rotation period, $\tau/P_{\rm rot}$, from random distributions which are listed in Table~\ref{tab:LC_inputs}. Each spot grows exponentially, peaks in size at a specific time and size, and then decays again exponentially. The times at which the spot sizes peak were drawn at random times from $-T$ to $2T$. This is to avoid a situation where, particularly for long lifetimes, the spot coverage is systematically lower at the beginning and end of the simulation than in the middle. The spots are assumed to be perfectly dark (contrast of 1 with respect to the stellar photosphere) and their peak sizes $A_{\rm max}$ range from 0.1 to 1\% of the visible hemisphere. The spots are randomly distributed on the stellar surface (uniform distribution in longitude, and in $\sin(\theta)$ where $\theta$ is spot latitude). This is not particularly realistic, but limiting the spots to specific latitude bands (as done for example in \citealt{2015MNRAS.450.3211A}) would not significantly alter the results of the present study, because the stars are seen equator-on. After reaching its peak size, each spot decays according to an exponential with half-life $\tau$. Following \citet{2015MNRAS.450.3211A}, the spots grow 5 times more rapidly than they decay (though we also performed some simulations with equal growth and decay time, to see if this has an effect on the recovery of the spot lifetime, as discussed in Section~\ref{sec:LC_evol}). We then simulate light curves using the analytic spot model of \citet{1987ApJ...320..756D}, using a linear limb-darkening law with $u=0.5$. Only idealised, noise-free data were considered in this study, so the simulated fluxes have no associated uncertainties. 

\begin{table}
\centering
  \caption{Stellar model input parameter distributions.    \label{tab:LC_inputs}}  
  \begin{tabular}{lc}
  \hline
  \hline
	Parameter			& parameter range (uniform distribution)\\
	\hline
	Average spot number, $N_{\rm Spot}$ & $0< \log_{10}(N_{\rm Spot}) < 2$ \\
    Stellar rotation period, $P_{\rm rot}$ & $0 < \log_{10}(P_{\rm rot}) < 2$\\
    Spot decay timescale, $\tau$ & $1\,P_{\rm rot} < \tau < 10\,P_{\rm rot}$\\
    Spot longitude, $\phi$ & $0 < \phi < 2 \pi$ \\
    Spot latitude, $\theta$ & $0< \sin(\theta)<1$\\
    Spot maximum area, $A_{\rm max}$ & $-3 \leq N_{\rm Spot} \,\log_{10} (A_{\rm max}) < -2$\\
    \hline
  \end{tabular}
\end{table}

\subsection{{Fitting the light curves using a GP}}
\label{sec:solve_the_GP}

When analysing a light curve simulated using a spot-model, the GP is no longer the `correct' model. Although the QP and QPC GP kernels described in Section \ref{sec:theGP} are very flexible, samples from GPs cannot reproduce the simulated light curves exactly (to machine precision). To account for this, we include a jitter term when modelling the light curves with the QP and QPC GPs. This jitter term absorbs any deviations between the simulated observations and the GP model. In practice, a jitter term is included by adding a constant value $J$ to the diagonal of the covariance matrix. We fit for $J$ alongside the other hyperparameters, resulting in more robust uncertainties on the latter.  The full set of parameters and GP hyperparameters that are solved for, and their associated prior distributions are given in Table \ref{tab:GP_priors}.

Initial guesses for each hyperparameter were defined as follows. The GP period $P$ was set to the stellar rotation period $P_{\mathrm rot}$, and the GP length scale $l$ to the spot decay timescale $\tau$. The GP mean and variance $A$ were set to the sample mean and variance of the simulated light curve. Since there is no single parameter of our stellar spot model that relates directly to $\Gamma$, we adopt an initial guess of $\Gamma=1$ across all models. This value marks the transition between high and low harmonic complexity cases \citep{PyanetiII}, and was chosen because it allows easy exploration of parameter space in either direction.
 
\begin{table}
\centering
  \caption{Priors placed on the parameters and QP GP hyperparameters for the MCMC fit of the model time series data  \label{tab:GP_priors}.}
  \begin{tabular}{ll}
  \hline
  \hline
	Input (hyper)parameter			& Prior$^{(\mathrm{a})}$\\
	\hline
	$\ln(A)$            & $\mathcal{U}[-30,30]$\\
	$\ln (P)$           & $\mathcal{U}[\log ({\rm max}[2T_{\rm min}, 1]),T_{\rm max}]^{(\mathrm{b})}$ \\
    $\Gamma$            & $\mathcal{U}[0,15]$\\
    $\ln (2l^2)$        & $\mathcal{U}[-30,30]$\\
    Mean                & $\mathcal{U}[-15,15]$\\
    $\ln ({\rm J})$& $\mathcal{U}[-20,20]$ \\
    \hline
  \end{tabular}
  \begin{tablenotes}\footnotesize
  \item \textit{Note} -- $^{(\rm{a})}$ $\mathcal{U}[a,b]$ refers to uniform priors between $a$ and $b$.
  $^{(\rm{b})}$  $T_{\rm max}$ and $T_{\rm min}$ refer to the maximum and minimum length of time between data points within a given time series. 
    \end{tablenotes}
\end{table}

The optimum GP hyperparameters were found using Markov Chain Monte Carlo (MCMC) with the python package {\sc emcee} \citep{emcee}. The MCMC was run with 80 walks for to a maximum of 10000 steps, unless the convergence was reached sooner. MCMC analyses struggle with `perfect', error-free data, so to help with convergence we specify a small uncertainty of $10^{-4}$ for the simulated data. The convergence of the MCMC walkers was checked and measured using the same method outlined in Section \ref{subsec:QPGPStability}. For the solutions that met this convergence criteria, a `burn-in' of twice the maximum chain auto-correlation length was removed, and the chains were thinned by half the minimum auto-correlation length. 

Given their greater complexity compared to the GP sample curves, the posterior distribution of the GP analysis of the light curves contains more local minima in which walkers can get 'lost'. We mitigate this in two ways: Firstly we set the walkers to follow a weighted combination of differential evolution (80\% probability) and differential evolution snooker (20\% probability) moves, as this is suited for multi-modal posterior distributions. Secondly, we implement a three stage burn in: the MCMC was run for 300 iterations and then reinitialised from the highest probability solution, repeating twice before the main MCMC run. If by the end of the main run there was not a converged solution for any of the reasons outlined above, a `burn in' of 5000 steps was removed form the start of the chains, the same test for convergence run again. If a solution passed, the same procedure for thinning and removing the `burn-in' was used as for the MCMC solutions that converged normally. If the solution still failed convergence, then no further burn-in or thinning was applied and a flag allocated to note it as a non-converged solution. 

\subsection{Results}
For all converged MCMC solutions, the final value was taken as the median value of the posterior distribution, and we estimate an uncertainty by calculating the the difference between the 50th and 16th percentile for the lower bound, and the 84th and 50th percentile for the upper. Any non-converged solutions were excluded from the analysis below. See Appendix \ref{sec:appA} for further discussion on the handful of model light curves for which the QP GP did not converge to a single set of solutions.

\subsubsection{Stellar rotation period and QP GP period}
\begin{figure}
	\centering
	\includegraphics[width=0.5\textwidth,trim={0 0cm 0 0cm},clip]{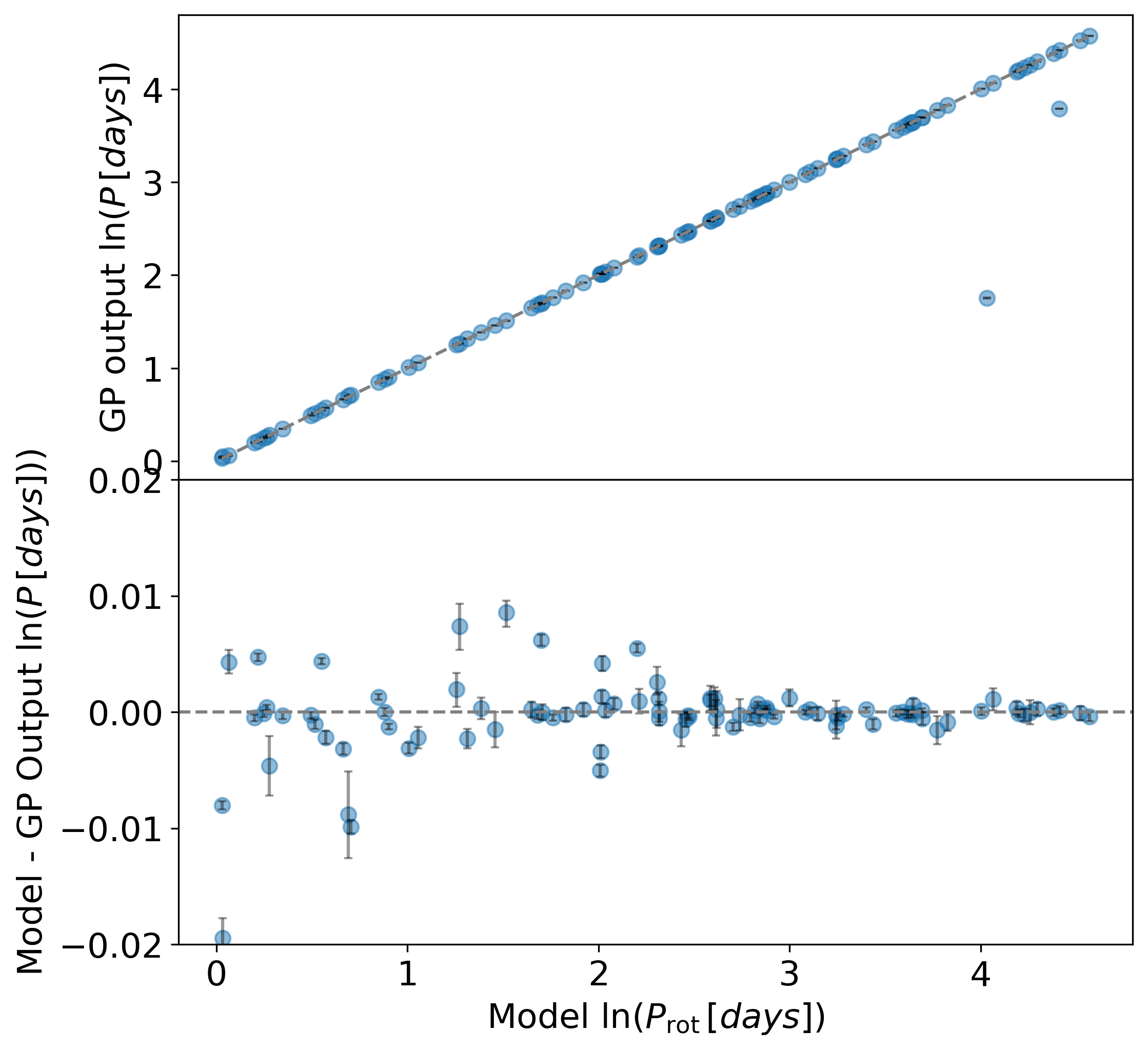}
    \caption{{The top panel compares the input stellar rotation period to the output QP GP period (both in natural log) for the model light curve data, with the unity relation indicated by the grey dashed line. The bottom panel shows the residuals to the unity relation, with axes scaled to exclude the larger outliers so as to more clearly show the scatter around the zero (grey dashed line).}}
    \label{fig:QPGP_LC_Period}
\end{figure}

A comparison between the stellar model rotation periods and the output QP GP period is shown in Figure \ref{fig:QPGP_LC_Period}. As seen in the top panel, the QP GP recovers the stellar rotation period for  all but one of the model light cures. For the two outlier points, the recovered period values are approximately a half and a tenth of the input model stellar rotation periods, owing to the particular distribution and evolution of the spots creating signals at those periods. 

The bottom panel shows the residuals of the GP period from the input model rotation period, excluding the outlier point to better see the scatter about zero. This scatter around zero (perfect recovery of the rotation period) is evenly distributed above and below, and increases for shorter rotation periods. This increased scatter is due to the finite sampling, with the longer periods being better sampled by the two points per day in our model light curve, and hence are better recovered by the GP.

\subsubsection{Stellar spot evolution time, $\tau$, and QP GP evolution time scale, $l$}
\label{sec:LC_evol}

\begin{figure}
	\centering
	\includegraphics[width=0.5\textwidth,trim={0 0cm 0 0cm},clip]{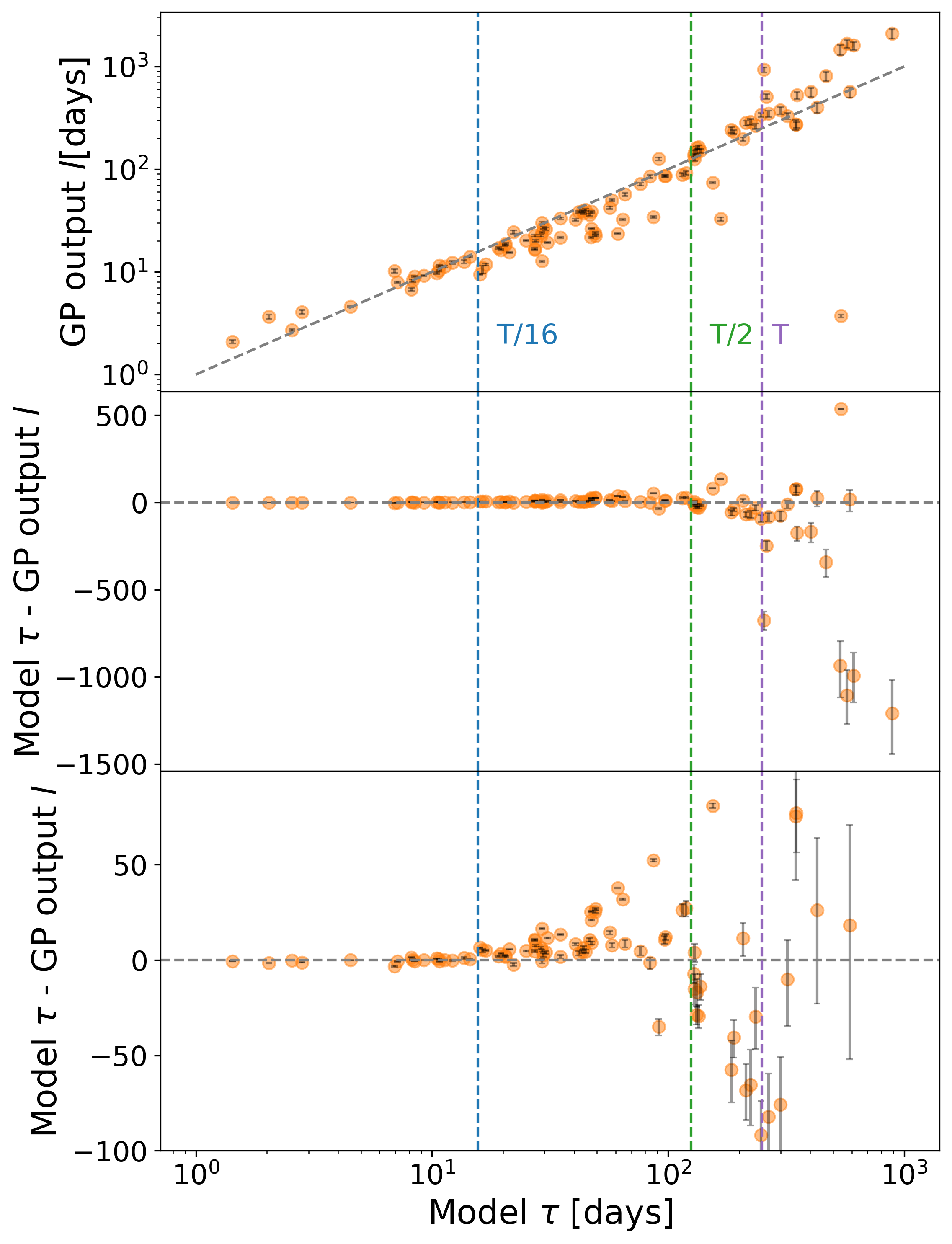}
    \caption{{This figure compares the input stellar model spot evolution time, $\tau$, to the solutions QP GP evolution time scale, $l$, for the set of model light curves (top panel), the residuals (middle), and the residuals with y-axis adjusted to highlight features at smaller $\tau$ (bottom). The grey dashed line indicates the unity relation, and the vertical dashed lines indicate the total (purple, T), half (green, T/2) and sixteenth (blue,T/16) time span of the data}. }
    \label{fig:LCEvolTimeVsLengthScale}
\end{figure}

Figure \ref{fig:LCEvolTimeVsLengthScale} compares spot evolution time scale, $\tau$, of the model stellar light curves with the QP GP length scale, $l$. The grey dashed line highlights the 1:1 correlation,  the purple vertical dashed line shows the total time span, T, of the model light curves of 250 days, and the green and blue vertical dashed lines indicate a half and sixteenth of the time span, respectively. 

These results show a strong correlation between the spot evolution timescale and the QP GP length scale hyperparameter. In the case of the specific way we have defined the spot behaviour in our stellar models, this is nearly a 1:1 relation, particularly at smaller values of $\tau$. The deviation from 1:1 recovery increases with increasing timescale $\tau$ as the values of $\tau$ approach the total time-span of the data, $T$. The QP GP length scale underestimates $\tau$ for values of $\tau$ between approximately T/16 and T/2, past which the GP overestimates $\tau$ with increasing scatter beyond $\tau$ greater than the time span of the data. 

\begin{figure}
	\centering
	\includegraphics[width=0.5\textwidth]{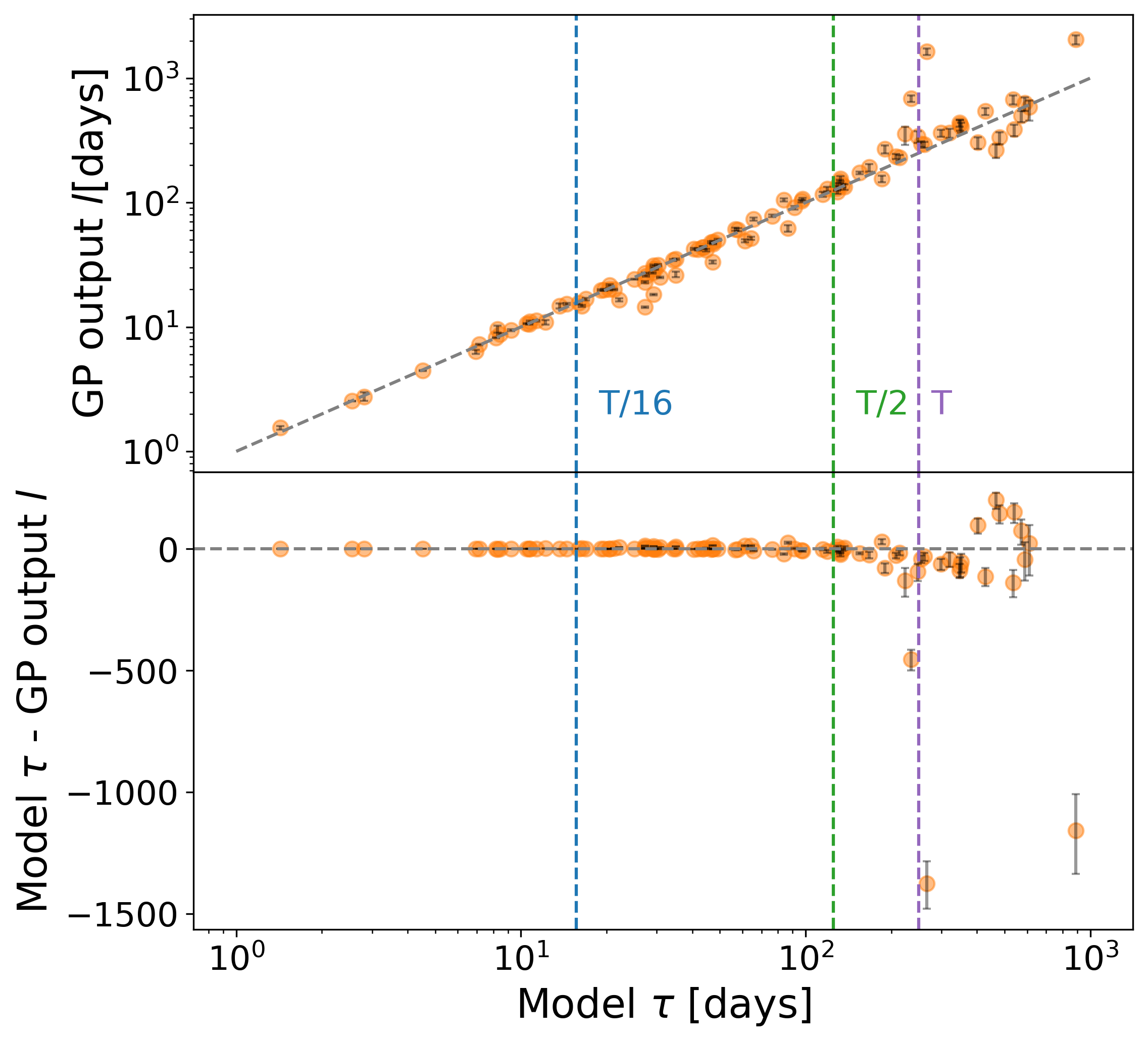}
    \caption{{Same as Figure \ref{fig:LCEvolTimeVsLengthScale} (direct comparison of values in the top panel, and residuals from the unity relation in the bottom panel), but for stellar models with equal spot emergence and decay times}}
    \label{fig:LC1EvolTimeVsLengthScale}
\end{figure}

In the stellar models used for this work, the spots emerge on the stellar surface five times faster than they decay, a choice motivated by observations of spot emergence and decay times on the Sun. To see whether this has an impact on the recovered QP GP length scale, we generated another set of simulated light curves with equal emergence and decay times, and modelled them in the same way. The results are shown in Figure \ref{fig:LC1EvolTimeVsLengthScale}. Comparing this to Figure \ref{fig:LCEvolTimeVsLengthScale}, the scatter around the 1:1 line is smaller for equal emergence and decay times. Again we see increasing scatter with increasing $\tau$, but this only becomes marked for values of $\tau$ approximately equal to or greater than the total time span of the data set (purple 
dashed line). {This result is unsurprising: when spot emergence is much faster than decay, the light curves are still dominated by changes occurring on the spot decay time, but this is occasionally perturbed by the emergence of a large spot causing a significant change on a much more rapid timescale. This is more challenging for our GP model, which has a single evolution length scale, to reproduce.} 

\subsubsection{Stellar model parameters and QP GP harmonic complexity, $\Gamma$}
\begin{figure}
	\centering
	\includegraphics[width=\columnwidth]{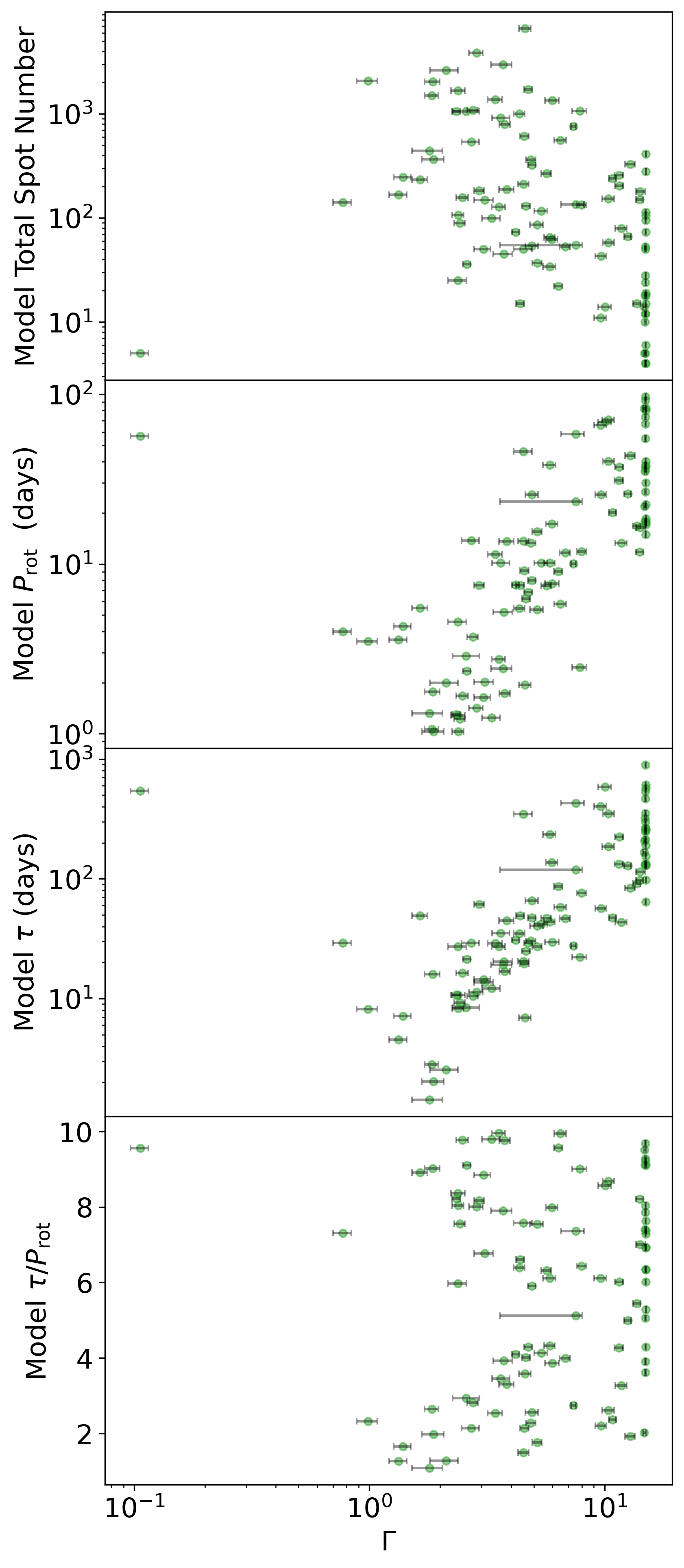}
    \caption{{Light curve MCMC solutions for the QP GP harmonic complexity, $\Gamma$, against the input stellar model parameters of total spot number, stellar rotation period, spot evolution timescale, and the ratio of the spot evolution timescale to the stellar rotation period. }}
    \label{fig:LCParsVsgamma}

\end{figure}

There is no intuitive physical interpretation for the harmonic complexity QP GP hyperparameter, so we examine how this hyperparameter behaves across all the input stellar model properties. Figure \ref{fig:LCParsVsgamma} shows the solutions for the QP GP $\Gamma$ with total spot number, stellar rotation period, spot evolution timescale, and the ratio of spot evolution time to the stellar rotation period. We calculate Pearson product-moment correlation coefficients of -0.36 with total spot number, 0.66 with stellar rotation period, 0.62 with spot evolution, and 0.12 with the ratio of spot evolution time to  stellar rotation period.

Since the $\Gamma$ term governs how complex the curve is \textit{within} one rotation period, the correlation with stellar rotation period results from longer period light curves being better sampled within one period, and so more complex behaviour can be resolved, giving a higher value for gamma. 

The correlation with evolution time, however, is spurious, as demonstrated by the bottom panel. By dividing through by the period, we remove all correlation of the evolution timescale with Gamma: at any given period, there is no correlation between $\tau$ and $\Gamma$. 

The slight, though not significant, correlation with total spot number is likely due to the difference in distribution of spots \citep[see e.g.][]{2015MNRAS.450.3211A}, which occurs when there are very few compared to many spots on the stellar disk. Given the uniform distribution imposed on the spot latitudes across all stellar models, models with a large number of spots will have an equally uniform spot distribution, but asymmetries in spot distribution will occur when there are very few spots.

Another influence in the outcome of these results are the input values and prior distributions used. Unlike the other hyperparameters that have initial values derived from the properties of the individual light curves, $\Gamma$ is given a starting value of 1 for all light curves. The walkers thus have further to explore to find an optimum solution, and may become `stuck' in a local preferred solution instead of finding the global optimum solution. However, as explained in Section \ref{sec:solve_the_GP}, this is mitigated by the multiple burn-in stages and setting a high maximum number of iterations to allow time for the walkers to fin the best solution. The clustering of points near $\Gamma=15$ is a result of the walkers reaching the upper bound of the uniform prior. Increasing this bound, however, does not change this behaviour, as it results from the GP trying to fit every small feature of the data that are due to the high-cadence sampling. This demonstrates the need for having a reasonable upper bound on $\Gamma$ so as to prevent over-fitting the data.

\subsection{{QPC versus QP GP analyses of light curves}}

{The same analysis was repeated with the QPC kernel, with the same MCMC framework and convergence criteria. There were two non-converged solutions, one of which had also failed to converge in the QP case (see Appendix \ref{sec:appA} for further discussion). When comparing the two analyses, all models that had a non-converged solution in either QP or QPC GP analyses were removed.} 

\begin{figure}
	\centering
	\includegraphics[width=0.5\textwidth,trim={0 0cm 0 0cm},clip]{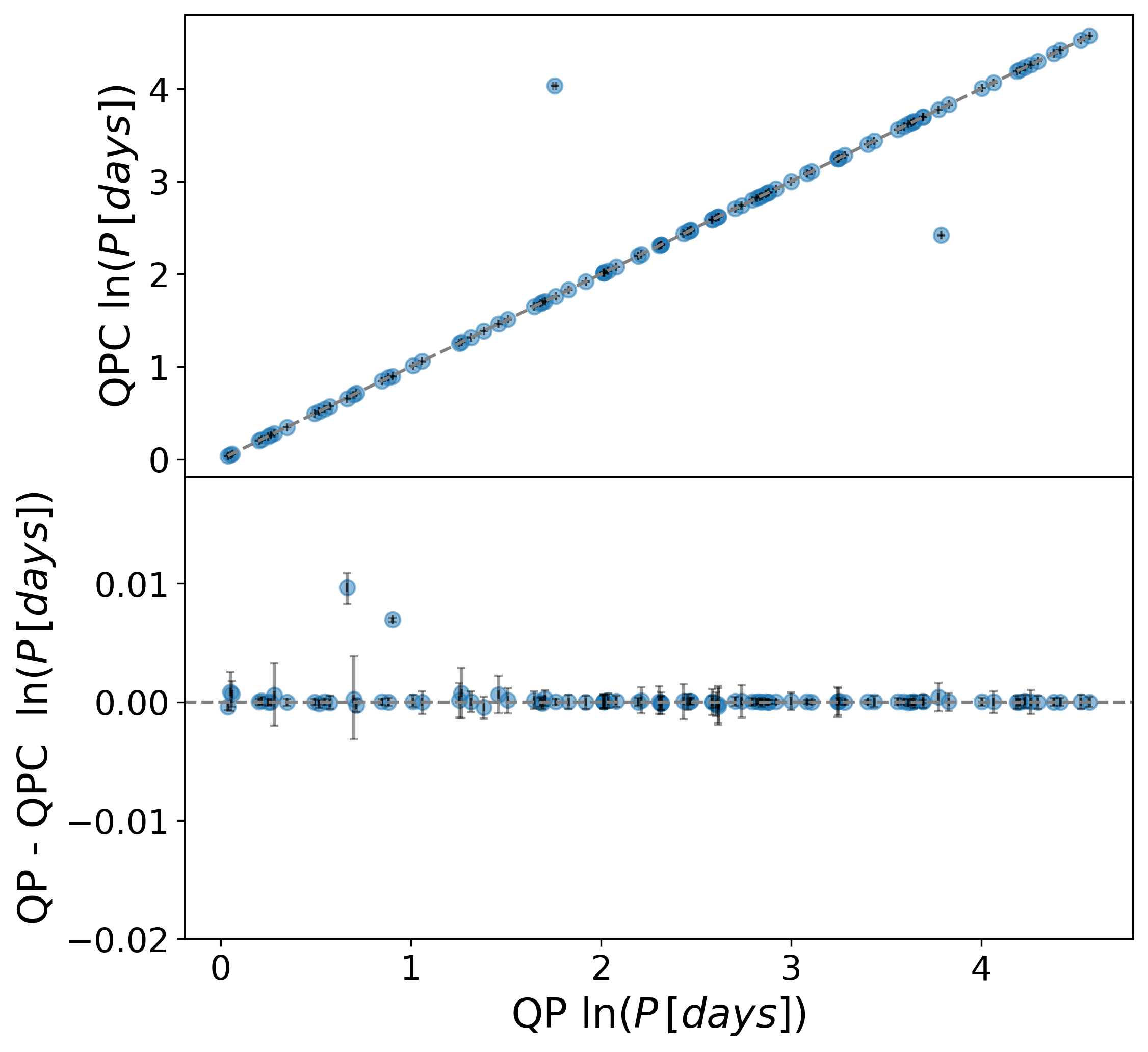}
    \caption{{This figure compares the QP and QPC solution for $\ln({\rm Period})$. Figure description is same as Figure \ref{fig:QPGP_LC_Period}}.}
    \label{fig:QPvQPC_lnP}
\end{figure}

\begin{figure}
	\centering
	\includegraphics[width=0.5\textwidth,trim={0 0cm 0 0cm},clip]{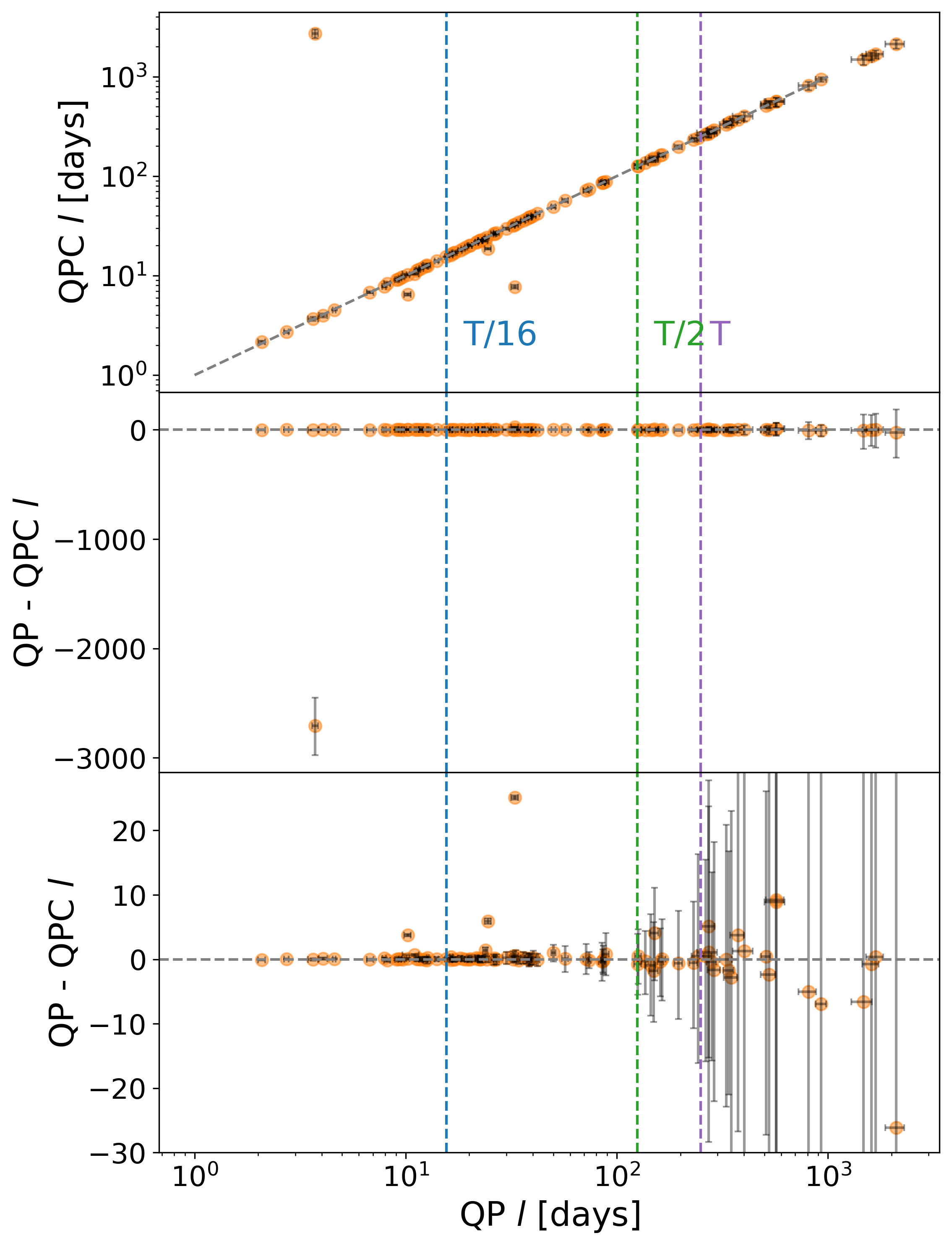}
    \caption{{This figure compares the solutions of the QP and QPC GP evolution time scale, $l$. Figure description is same as Figure \ref{fig:LCEvolTimeVsLengthScale}.}}
    \label{fig:QPvQPC_l}
\end{figure}

\begin{figure}
	\centering
	\includegraphics[width=0.5\textwidth,trim={0 0cm 0 0cm},clip]{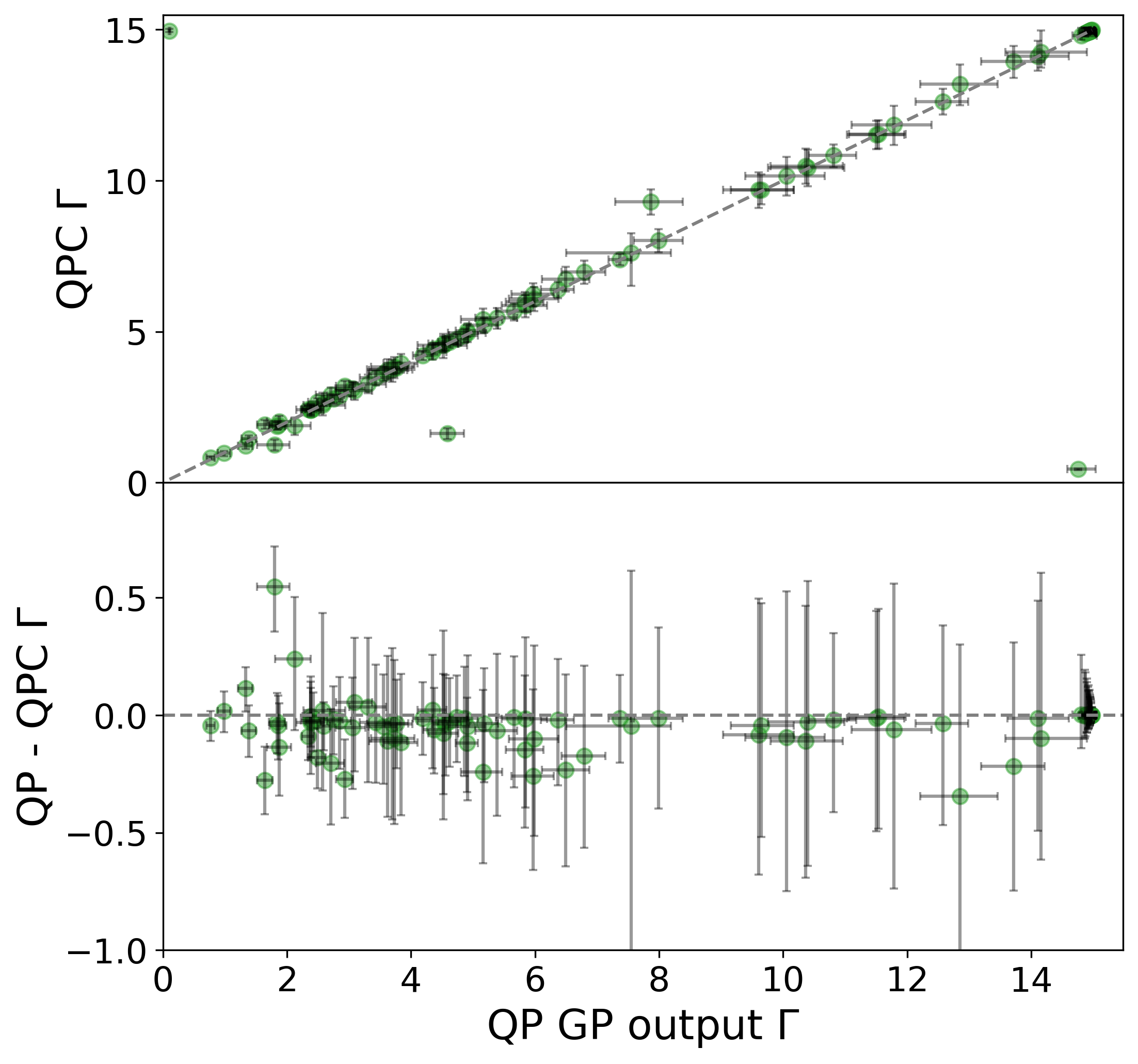}
    \caption{{Top panel of this figure compares the QP and QPC kernel solutions for the harmonic complexity, $\Gamma$, with grey dashed line indicating the unity relation. Bottom panel shows the residuals around the one-to-one relation, with y-axis scaled to exclude the 4 largest outliers.}}
    \label{fig:QPvQPC_gamma}
\end{figure}

{Figures \ref{fig:QPvQPC_lnP}, \ref{fig:QPvQPC_l} and \ref{fig:QPvQPC_gamma} show the comparison between the hyperparameters that are commmon to the QP and QPC kernels. Across all three hyperparameters the values QP and QPC values are almost identical. }

{For period, the agreement between the QP and QPC solutions is excellent, even at shorter period where scatter between the input rotation period and output QP GP period is higher. Additionally, there are one fewer outliers in the QPC kernel results. The QPC kernel is expected to better handle P/2 aliases in the data, but this is not seen in this case. The outlier in the QP case at $\sim 0.5 P$ remains an outlier here, and has been underestimated ever further to $\sim 0.1$ of the rotation period. The outlier that was at $\sim 0.1P$ in the QP case, on the other hand, has converged to the input rotation value in the QPC case. The models with that are the largest outliers in period are also the largest outliers in $l$ and $\Gamma$.}

{The results for evolution timescale $l$ between the QP and QPC analyses are also in excellent agreement. They show increasing scatter at longer $l$, though all agree within the MCMC solution uncertainties. This scatter begins to increase at T/2, and increases significantly past T, as seen most clearly in bottom panel of Figure \ref{fig:QPvQPC_l}. }

{Figure \ref{fig:QPvQPC_gamma} demonstrates the excellent agreement between the QP and QPC solutions for $\Gamma$. In particular this is highlighted the lower panel of this figure, which has been scaled to exclude the largest residuals and better see this scatter about zero. Above $\Gamma \sim 2$ (and ignoring the points clustered at the edge of the uniform prior boundary values of 15), the scatter in the values is smaller than the MCMC uncertainties. Within those uncertainties, all residual points can be said to fall on zero, although all QPC points are systematically slightly larger than the QP values. }

\begin{figure}
	\centering
	\includegraphics[width=0.5\textwidth,trim={0 0cm 0 0cm},clip]{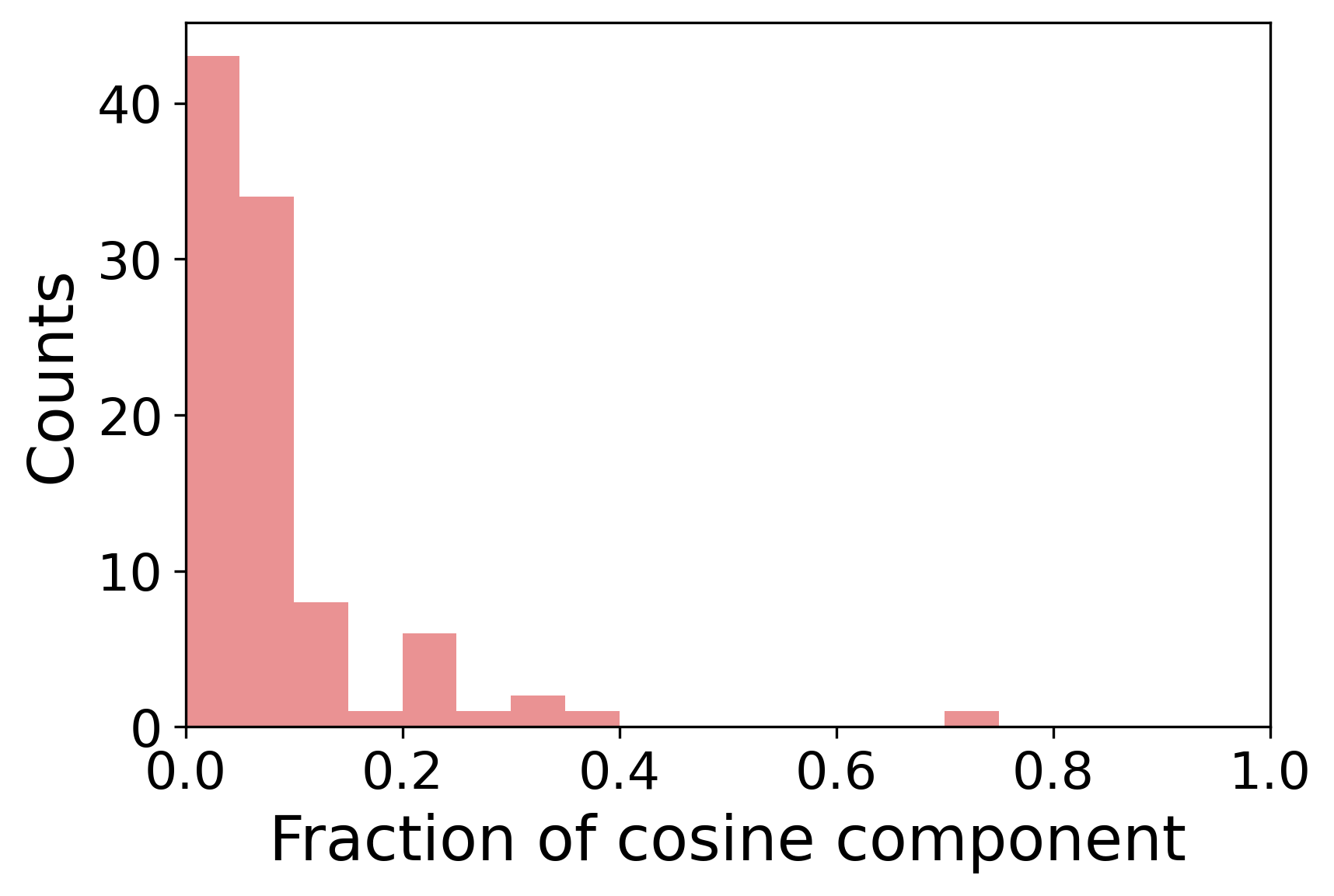}
    \caption{{This figure shows a histogram of the fraction of cosine component, $f$, values for the QPC GP for to our model light curves.}}
    \label{fig:QPC_frac_hist}
\end{figure}

{The near identical results seen between the QP and QPC kernels presented above can be explained by the distribution of $f$ values. As shown in Figure \ref{fig:QPC_frac_hist}, all but a few of the converged solution have solution for $f$ that are very small and tend toward zero. As such, these GPs are behaving more like the QP kernel, and hence giving near identical results for period, $l$ and $\Gamma$. This is surprising, as more than a third of our model curves have some signal at half the rotation period (seen as a peak in their auto-correlation function at P/2), and thus are expected to have values of $f$ greater than zero. It is possible that the cosine term in the QPC kernel only requires a small amplitude to have a significant effect, but our results may also indicate that QP GP alone handles signals at $P/2$ almost as effectively as the QPC kernel.}

\section{Applying the QP kernel to model RV time series}
\label{sec:GPofRV}

It is common practice to use photometric data to inform the priors for QP GP analyses of RV data, often used to mitigate stellar activity signals in the search for Keplarian signals from exoplanets. To explore how the hyperparameters of a QP GP fit to photometric data will compare to that of RV data, we use the same spot models used above to compute RV time-series using the following, very simple procedure  \citep{Aigrain2012}: at each time-point, we compute the local RV of the stellar surface at the centre of each spot, and multiply it by the spot's contribution to the light curve (in relative flux units). 

The resulting RV time-series have the same sampling and duration as the light curves. While they are not representative of real RV data sets, which are typically ground-based and have much sparser time-sampling, we use these RV simulations to explore the differences between the hyperparameters of the QP GP models trained on the photometric and RV time series. The impact of the difference in time sampling between space and ground based observations will be explored separately in Section~\ref{sec:RV_resamp}. 

The same MCMC procedure was used to find solutions for the hyperparameters of a QP GP applied to the RV time series, and the treatments of non-converged solutions was the same. The {two} stellar models that gave multi-modal (non-converged) solutions in the photometric data, had converged, single solutions in the RVs, and a handful of the stellar models that had converged solutions for the LC data had non-converged solutions here. To best compare the LC and RV results, all models that non-converged solutions in either analysis were removed. 

\subsection{RV vs light curve period}
\begin{figure}
	\centering
	\includegraphics[width=\columnwidth]{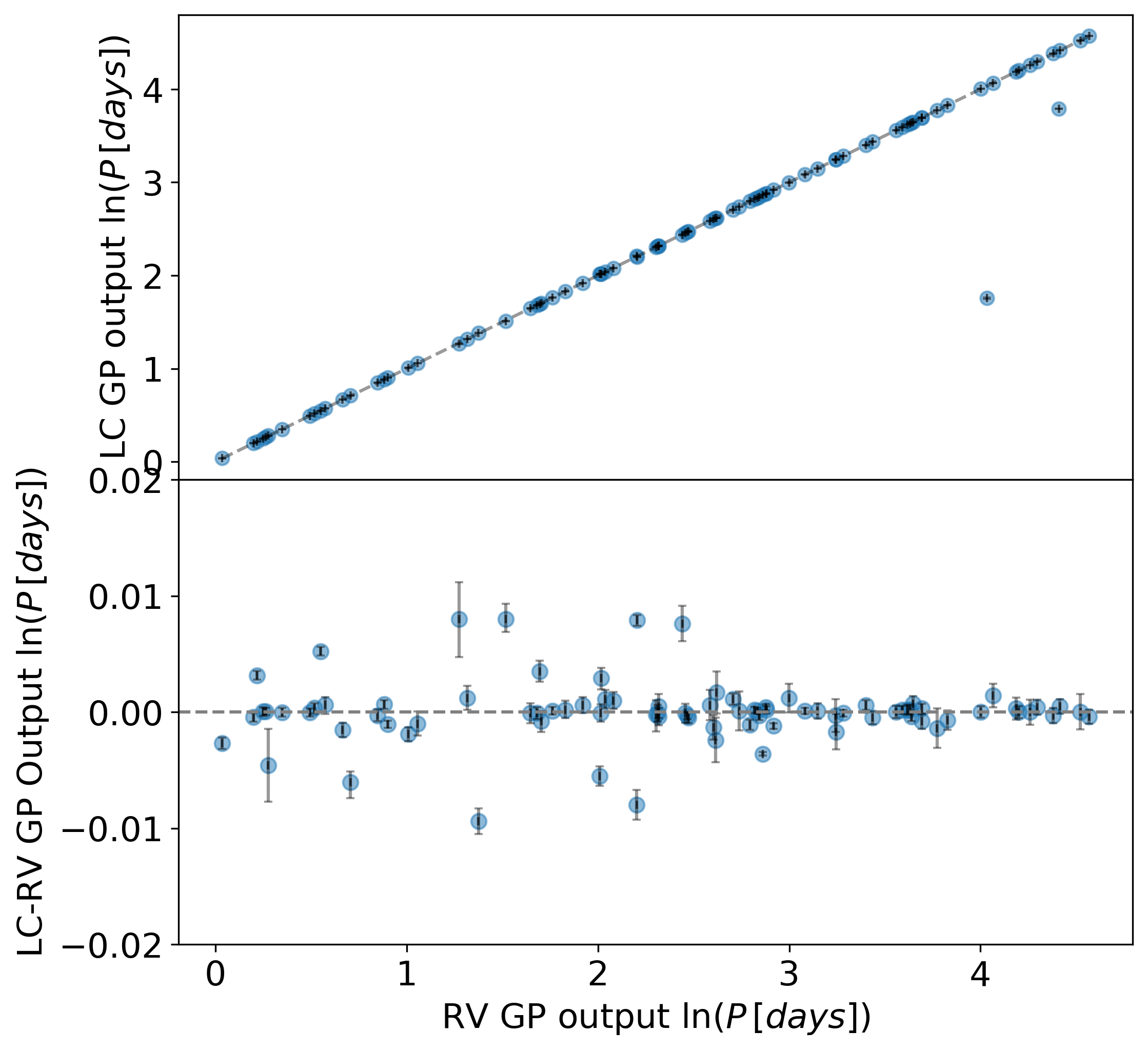}
    \caption{{The top panel compares the output QP GP period (in natural log) for the radial velocity (RV) data with the light curve (LC) data, with the unity relation indicated by the grey dashed line. The bottom panel shows the residuals to the unity relation, with axes scaled to exclude the larger outliers and more clearly show the the scatter around the zero (grey dashed line).} }
    \label{fig:LCvsRV_Period}
\end{figure}

The comparison of the solution to the QP GP period from light curve versus RV data is shown in figure \ref{fig:LCvsRV_Period}. The top panel plots the GP solution for the light curve against those of the RVs, with the grey dashed line showing the unity relation, and the bottom panel shows the residuals from this, excluding the outlier seen in the upper panel results from the underestimated period of one light curve QP GP solution. The period solutions for the RV QP GP shows no such outliers, and recover well the model stellar rotation periods. The lower panel show a greater scatter in the residuals for shorter periods, reflecting the scatter seen  in the light curve GP solutions. Overall, the light curve and RV QP GP solutions for the period are in excellent agreement, both reflecting the stellar rotation period. 

\subsection{RV versus light curve $l$}
\begin{figure}
	\centering
	\includegraphics[width=\columnwidth]{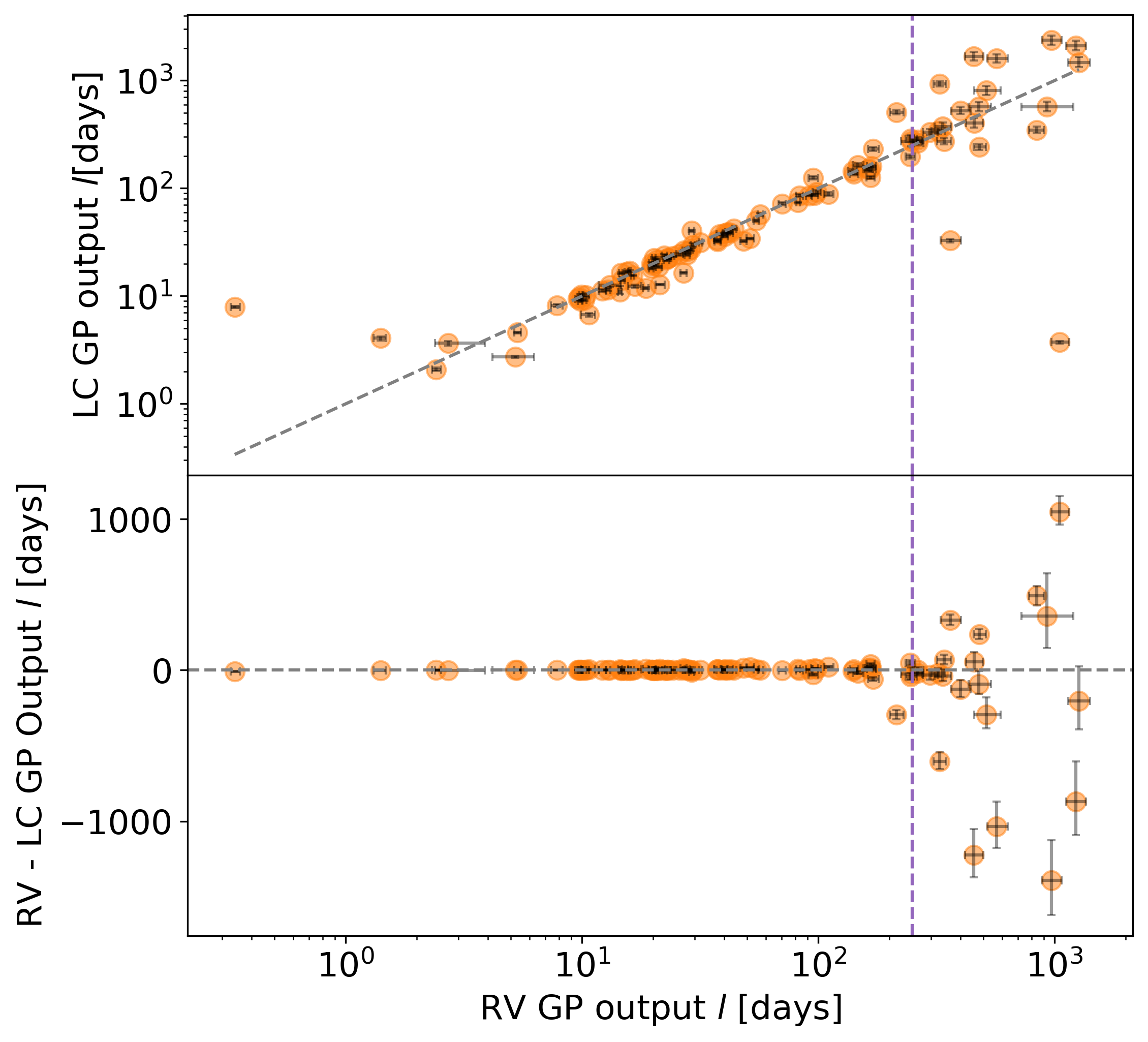}
    \caption{{This figure compares the solutions QP GP evolution time scale $l$ for model light curves (LC) with the solutions for the radial velocity (RV) data. The grey dashed line indicates the unity relation, and the purple vertical dashed lines indicate the total time span of the date of 250 days.}}
    \label{fig:LCvsRV_l}
\end{figure}

Figure \ref{fig:LCvsRV_l} compares the QP GP length scale solution from RV time series versus the solution from model light curve data, with the grey dashed line indicating the unity relation, and the vertical purple dashed line showing the total time span of the data of 250 days. The light curve and RV solutions are in good agreement, but differ more at larger length scale, particularly for those length scales greater than the time span of the data.

\subsection{RV versus light curve $\Gamma$}
\begin{figure}
	\includegraphics[width=\columnwidth]{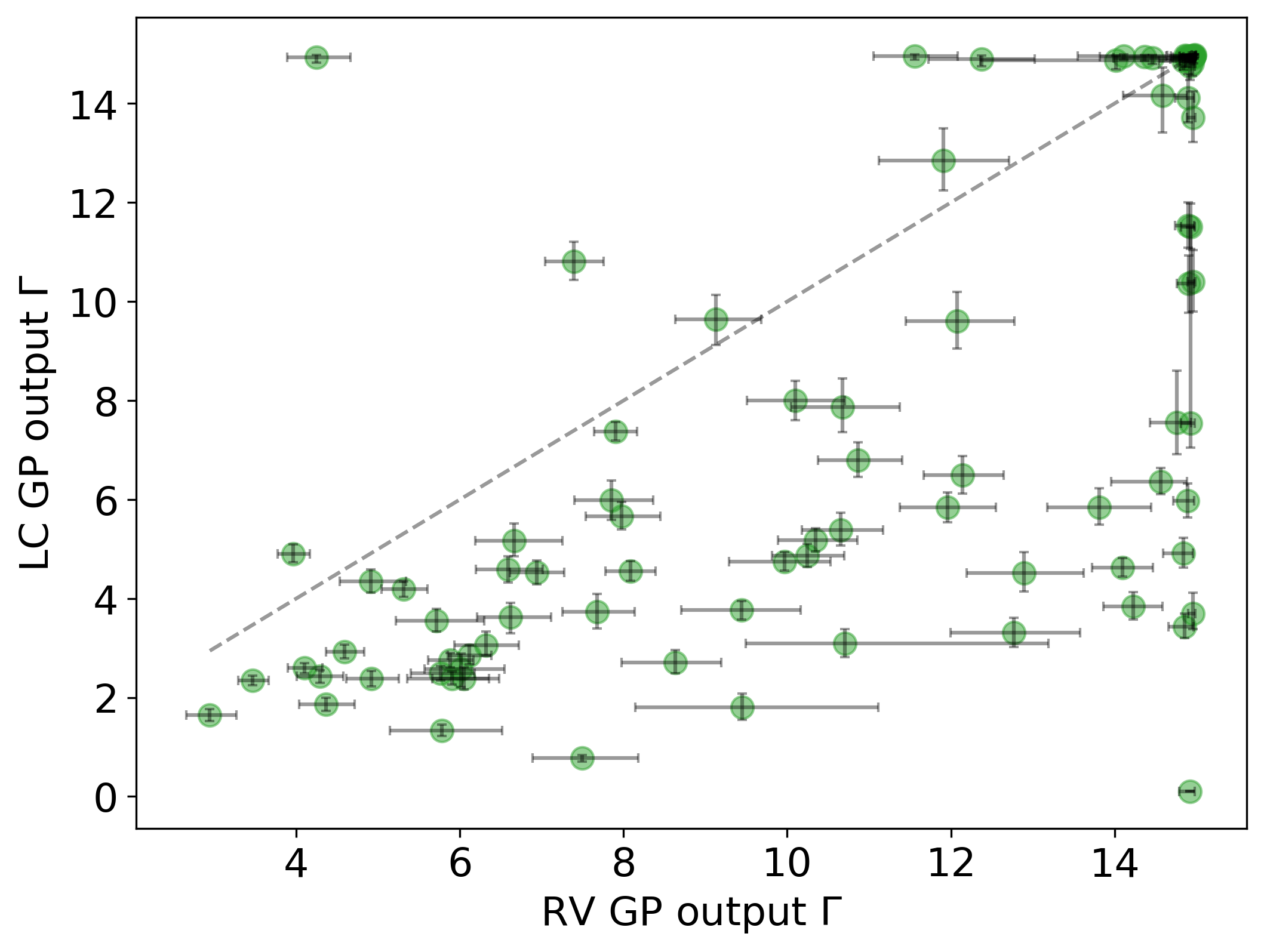}
    \caption{{The RV recovered harmonic complexity term ($\Gamma$) of the QP GP, versus that from the light curve (LC) data. The unity relation is shown by the grey dashed line.} }
    \label{fig:LCvsRV_gamma}
\end{figure}

Figure \ref{fig:LCvsRV_gamma} compares the RV and light curve solutions of the QP GP harmonic complexity, $\Gamma$, with the 1:1 relation shown as a grey dashed line. Unlike the period and the length scale, the RV and light curve QP GPs have very different solutions for $\Gamma$. The RV solution gives a consistently larger $\Gamma$ than the light curve data from the same stellar model. Further, the moderate correlations seen in the light curve solutions between $\Gamma$ and the stellar rotation period is reduced in the solutions for RVs, with a Pearson correlation coefficient of 0.5. There remains a slight, but weaker anti-correlation with total spot number of -0.29. 

The lack of agreement between the harmonic complexity term between RV and photometric data has implications for the way we approach activity modelling in RV data sets (see Section \ref{sec:DiscConc}. 

\subsection{QPC versus QP GP analyses of RVs}
We repeat the GP analysis of the model RVs with the QPC kernel, again using the same MCMC framework, convergence criteria and treatment on non-converged solutions. Only one model had non-converged solutions, and this was a different model to those that didn't converge in any other GP analysis run (see Appendix \ref{sec:appA} for further discussion). 

\begin{figure}
	\centering
	\includegraphics[width=0.5\textwidth,trim={0 0cm 0 0cm},clip]{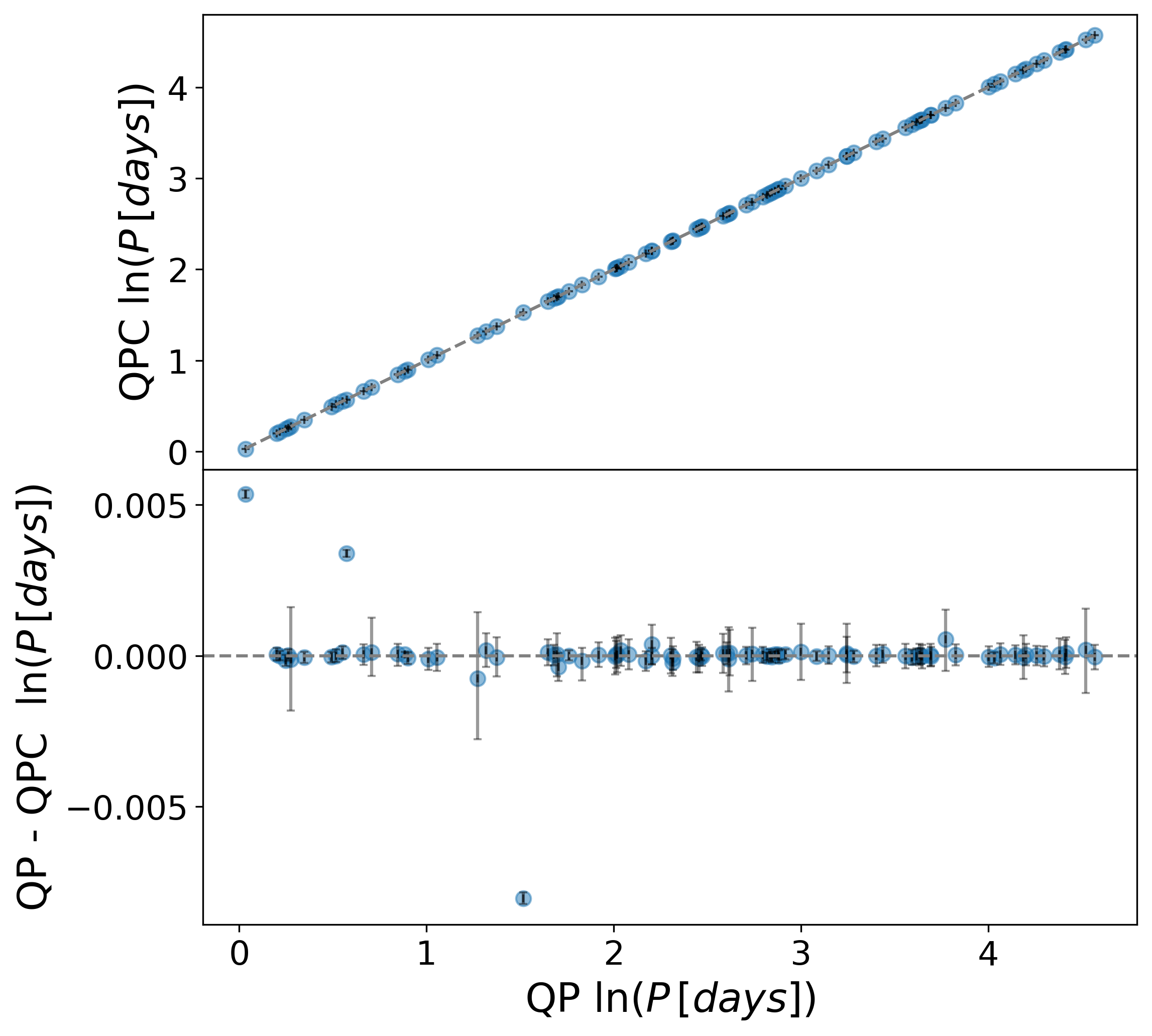}
    \caption{{This figure compares the QP and QPC solution for $\ln({\rm Period})$. Figure description is same as Figure \ref{fig:QPGP_LC_Period}}.}
    \label{fig:QPvQPCRV_lnP}
\end{figure}

\begin{figure}
	\centering
	\includegraphics[width=0.5\textwidth,trim={0 0cm 0 0cm},clip]{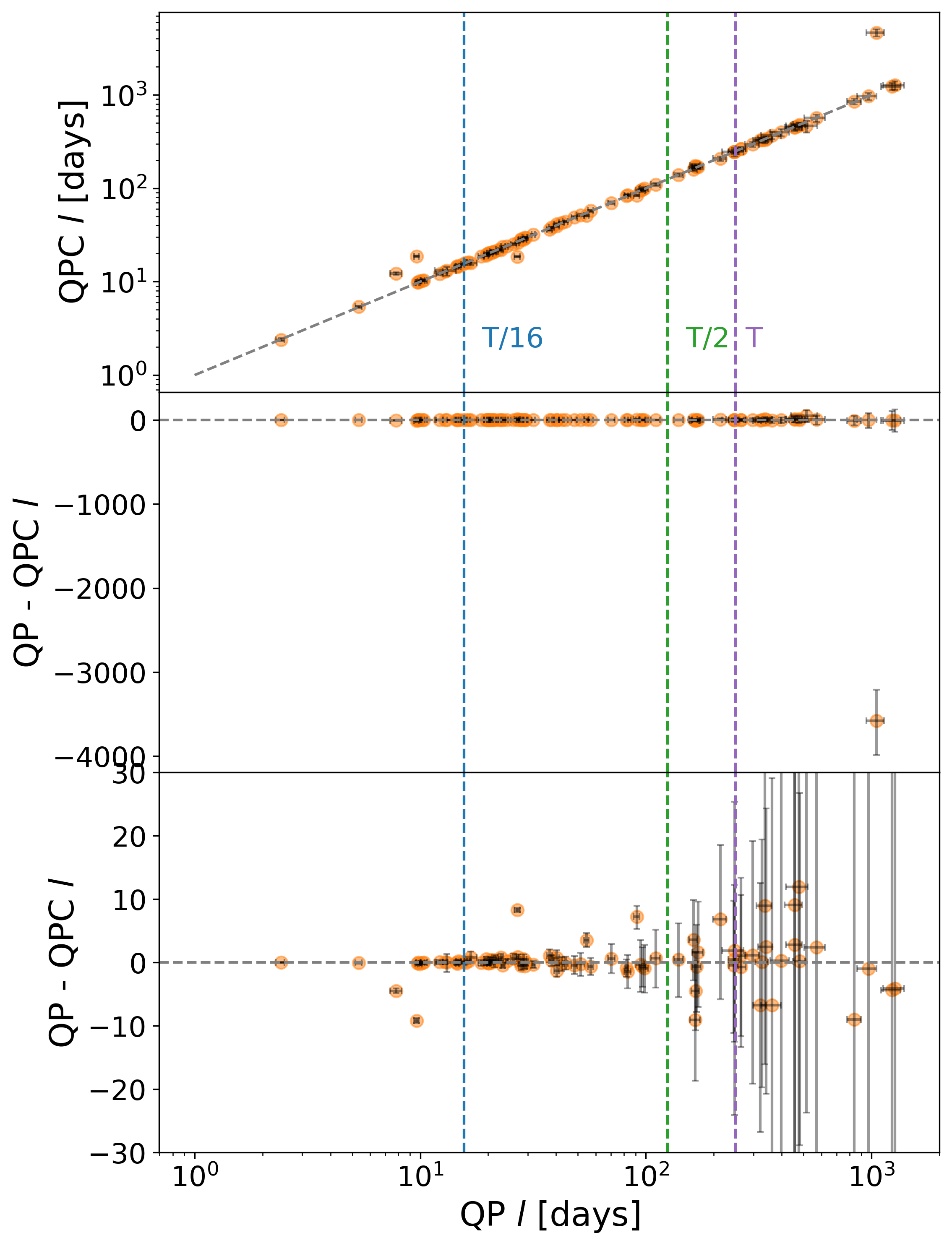}
    \caption{{This figure compares the solutions of the QP and QPC GP evolution time scale, $l$. Figure description is same as Figure \ref{fig:LCEvolTimeVsLengthScale}.}}
    \label{fig:QPvQPCRV_l}
\end{figure}

\begin{figure}
	\centering
	\includegraphics[width=0.5\textwidth,trim={0 0cm 0 0cm},clip]{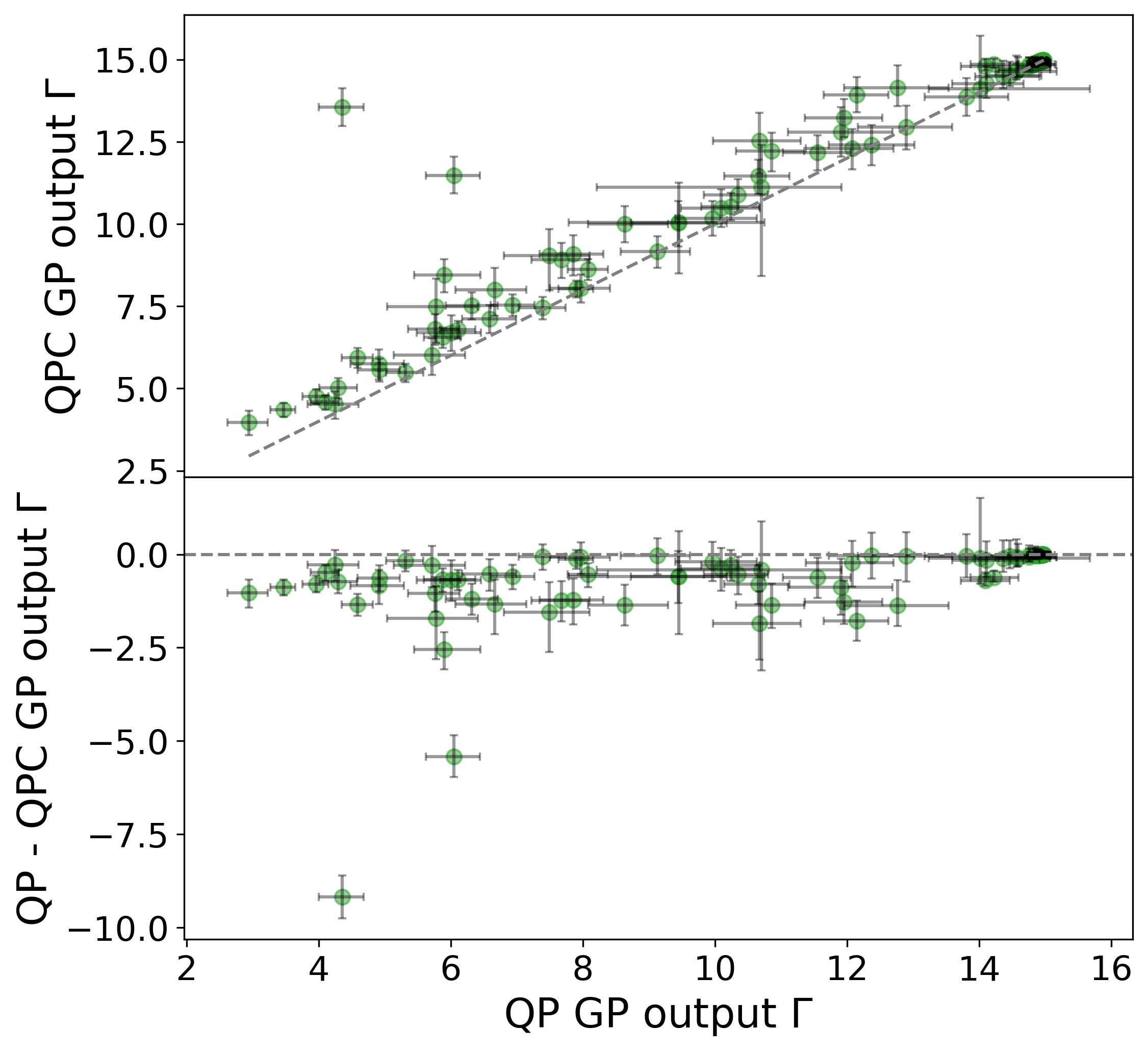}
    \caption{{This figure compares the QP and QPC GP solutions for $\Gamma$ for analysis of model RV data with the unity relation shown as a grey dashed line (top panel), and the residuals from the 1:1 relation with the grey line indicating zero residuals (bottom panel)}.}
    \label{fig:QPvQPCRV_gamma}
\end{figure}

A comparison of results between the QP and QPC kernels for model RV data are shown in figures \ref{fig:QPvQPCRV_lnP}, \ref{fig:QPvQPCRV_l} and \ref{fig:QPvQPCRV_gamma}. As was the case for the  analysis of the model light curves, the QPC solutions for period, evolution timescale and harmonic complexity are nearly identical to those of the QP kernel. 

Examining the residuals of the period comparison (lower panel of Figure \ref{fig:QPvQPCRV_lnP}), there are three notable outliers. These models are also the three largest outliers in $\Gamma$ but not in $l$. As in the light curve analysis, the scatter in the difference between the QP and QPC kernels is smaller than the uncertainty, and does not vary with period. 

Similarly with the evolution timescale, $l$, aside from some outliers, the scatter in the residuals about zero is smaller than the uncertainties, with the scatter and degree of uncertainty increasing for longer timescales. 

\begin{figure}
	\centering
	\includegraphics[width=0.5\textwidth,trim={0 0cm 0 0cm},clip]{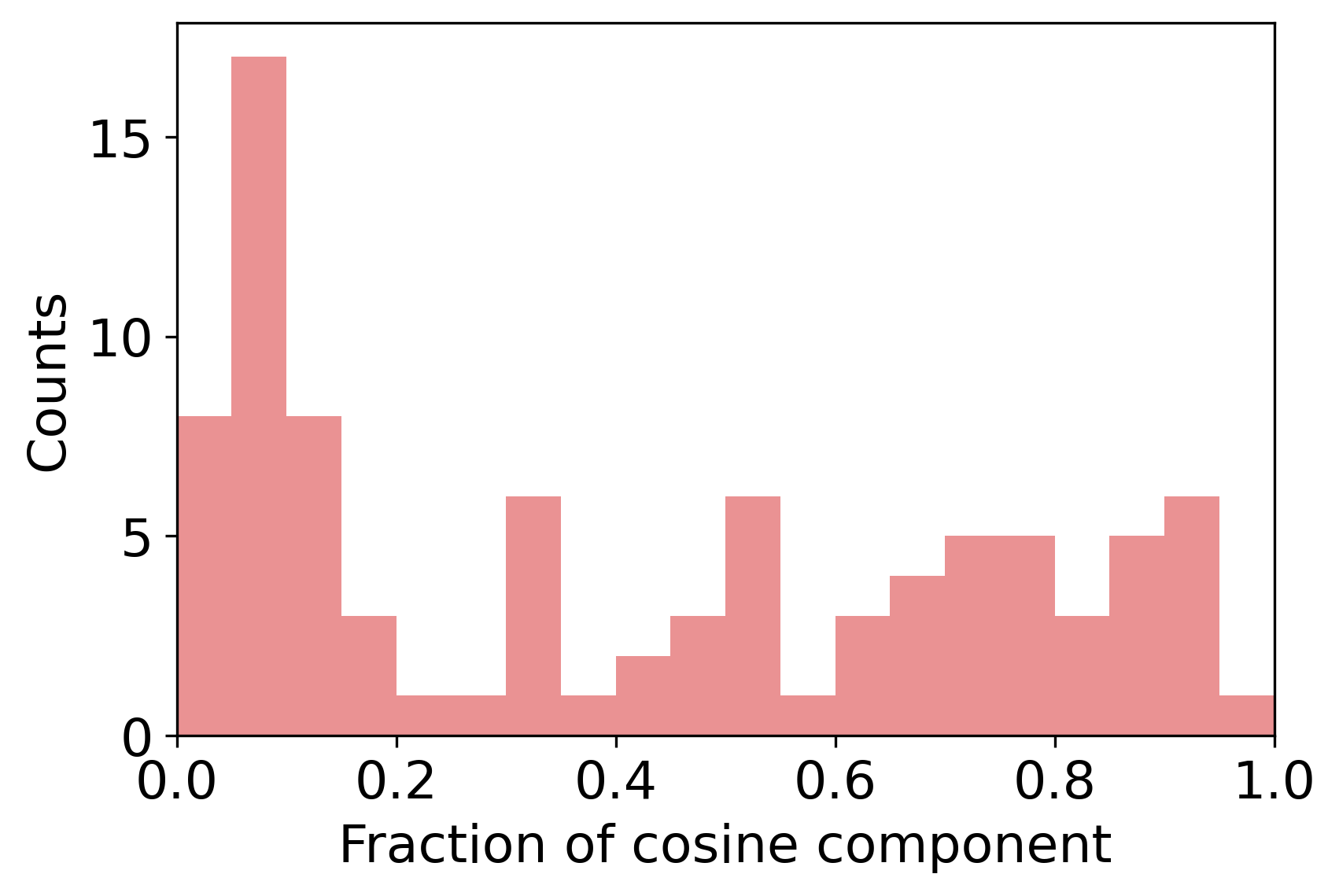}
    \caption{{This figure shows a histogram of the fraction of cosine component, $f$, values for the QPC GP for to our model light curves. }}
    \label{fig:QPC_RV_frac_hist}
\end{figure}

The harmonic complexity shows different behaviour here compared to the LC analysis, with the the QPC $\Gamma$ values being consistently higher than the QP solution for the same stellar model RV. This offset in the harmonic complexity can be explained by the higher fraction of cosine component $f$ in the QPC solutions of the model RVs. As seen in Figure \ref{fig:QPC_RV_frac_hist}, many more models had a values of $f$ above 0.1, and thus the cosine component was accounting for more of the intra-period signal, leaving the harmonic complexity term to fit to more of the subtle variations from point to point, increasing the complexity term overall - without the larger, simpler signal to fit to, the $\Gamma$ parameter was instead sensitive to the smaller-scale changes.

\section{Impact of RV time sampling}
\label{sec:RV_resamp}

RV observations are typically ground-based, and as such their time-sampling is generally much sparser than that of the simulated data sets analysed in the previous Section{, and have much higher uncertainties}. To test the application of QP GP regression to more realistic RV data, we {add noise to these data representative of the uncertainties expected at world-leading precision radial velocity facilities with a mean uncertainty of 20cm/s, and a standard deviation of 15cm/s to approximately reflect the night-to-night variations caused by different atmospheric conditions. We then} re-sample the model RV curves with the type of sampling expected form ground-based observatories by choosing a real star at random, and calculating its observability from a chosen observatory given seasonal and air mass constraints, as one would when planning a RV observing season. We then randomly selected $N$ observations within the observable times, over a time span of 250 days, with $N=250$, $100$, $50$ and $30$. While 30 to 50 points per season are typical of fairly intensive, present-day RV monitoring campaigns, 100 to 250 points per season corresponds to the planned sampling rates for future EPRV surveys aiming to detect Earth analogues, such as the Terra Hunting Experiment (THE, \citealt{2018MNRAS.479.2968H}). To ensure adequate sampling of the rotation periods for even the most rapidly-rotating simulated stars, we added a further constraint that the shortest gap between any two data points in a given time series should not be longer than the shortest rotation period in our simulations. This mimics what an observer would do when planning observations.

\begin{figure}
    \centering
    \includegraphics[width=0.90\columnwidth]{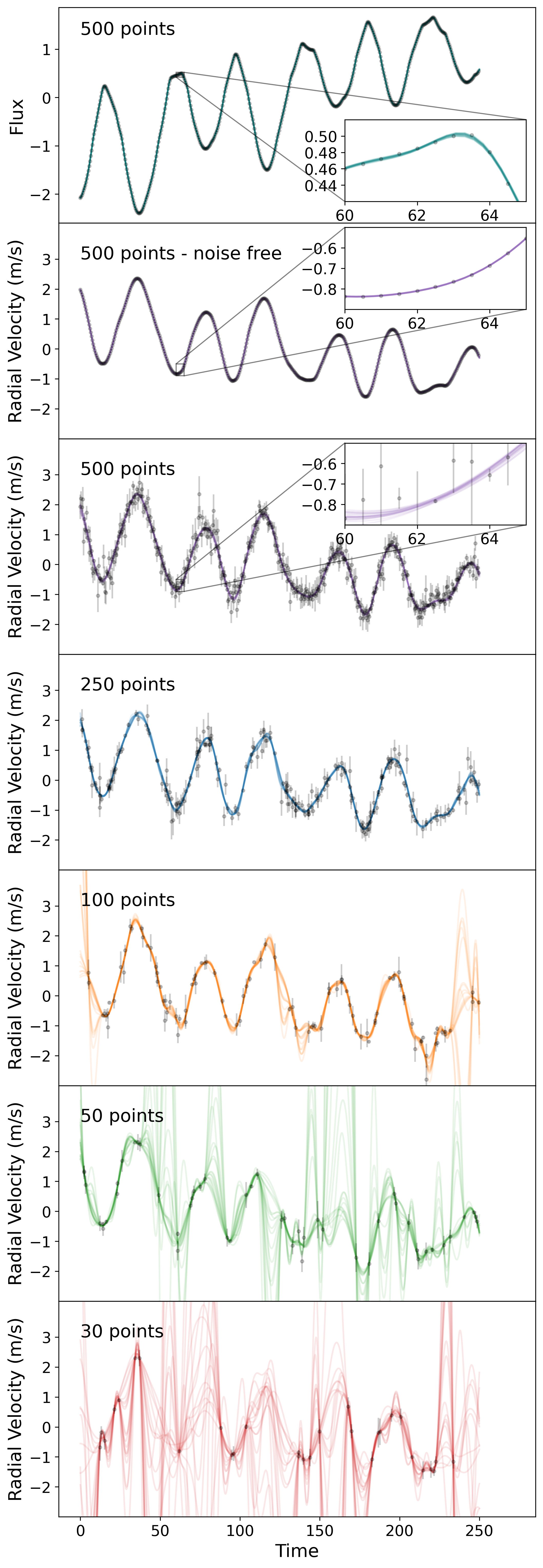}
    \caption{{This figure shows an example model light curve (top panel), its associated RV data, and the RV data sets with degraded sampling. The lines plotted in each are the QP GP model with hyperparameters drawn from the MCMC posteriors.} }
    \label{fig:RV_resamp_08}
\end{figure}

An example of this progressive degradation of the time-sampling {and increased errors} is shown for one of the stellar models in Figure \ref{fig:RV_resamp_08}, with the data points shown in grey. The re{-}sampled light curves we analysed in the same way as light curves and the original RV curves, using the same MCMC sampling and convergence criteria. As before, we exclude any non-converged solutions from our analysis. The lines in each panel of \ref{fig:RV_resamp_08} are samples from the QP GP conditioned on the data, with hyperparameters drawn from the MCMC posterior distribution. Larger scatter between the curves indicate that the GP model is less well constrained. The effect of reducing the number of observations by a factor up to 5 (down to $N=100$) appears negligible in these plots, but it becomes noticeable when $N=50$ or $30$. 

\begin{figure}
	\includegraphics[width=\columnwidth,trim={0 0cm 0 0cm},clip]{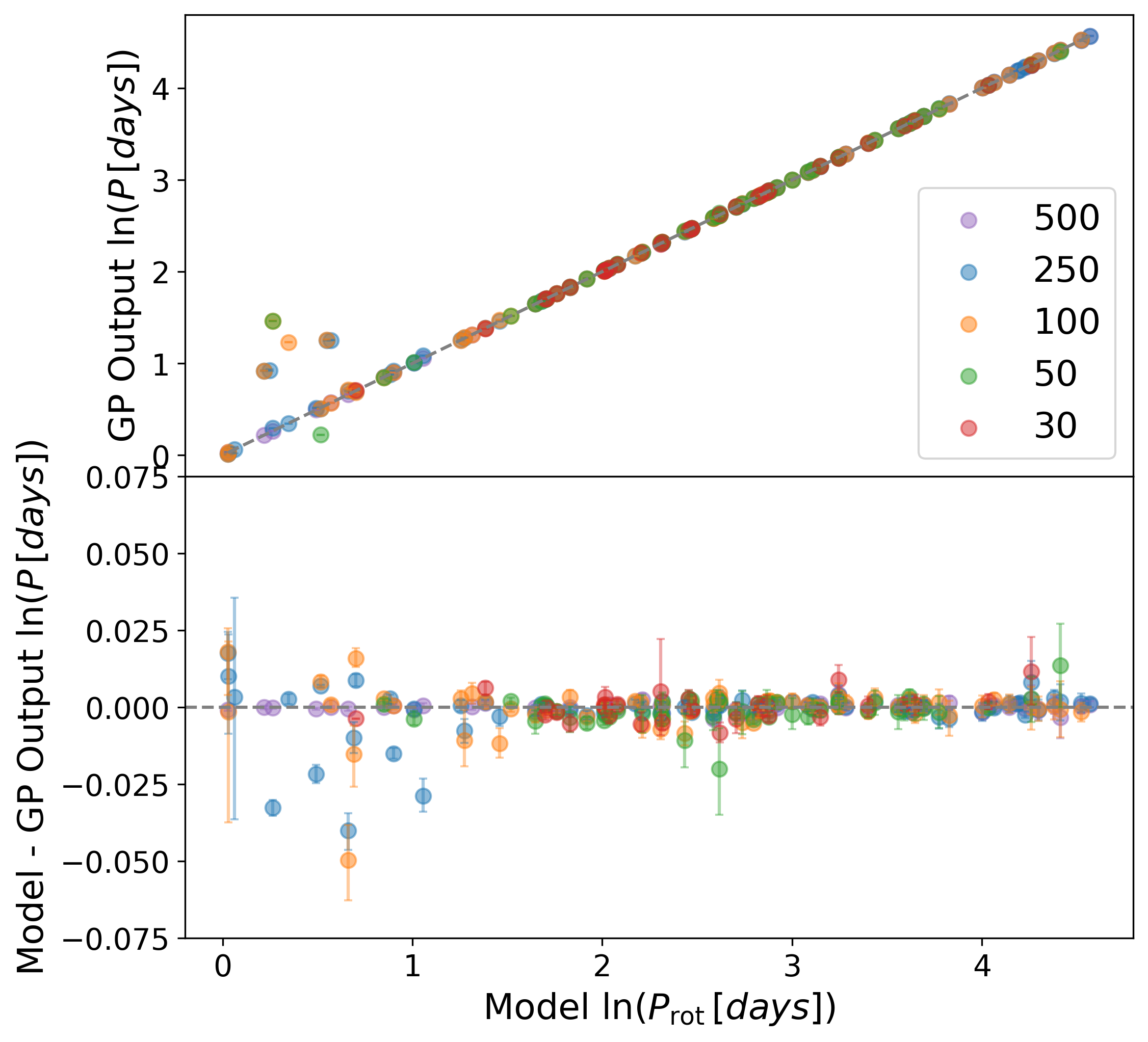}
    \caption{{The top panel compares the input stellar rotation period to the output QP GP period (both in natural log) for different different time samplings of the model RV data, with the unity relation indicated by the grey dashed line. The bottom panel shows the residuals to the unity relation, with axes scaled to exclude the larger outliers and more clearly show the the scatter around the zero (grey dashed line).} }
    \label{fig:RV_resamp_lnP}
\end{figure}

We then compare the input stellar rotation period and output GP period in Figure \ref{fig:RV_resamp_lnP}. In the bottom, which shows the residuals from a one-to-one correspondence, we have excluded the clear outliers visible in the top panel, and zoomed-in to better show the behaviour of the remaining data points. With decreased sampling, the scatter around the input value increases for smaller values of the stellar rotation period, as these periods are less well resolved by the data. For d$N =50$, only three cases with with rotations periods below 4 days ($\ln(P) < 1.38$) converged. For $N=30$ this number went down even further to one. This highlights the importance of sampling both the stellar rotation period and the orbital period(s) of any candidate planets adequately. For rapidly rotating stars, this necessarily translates into large numbers of observations per season unless the planet periods are also short.

\begin{figure}
	\includegraphics[width=\columnwidth,trim={0 0cm 0 0cm},clip]{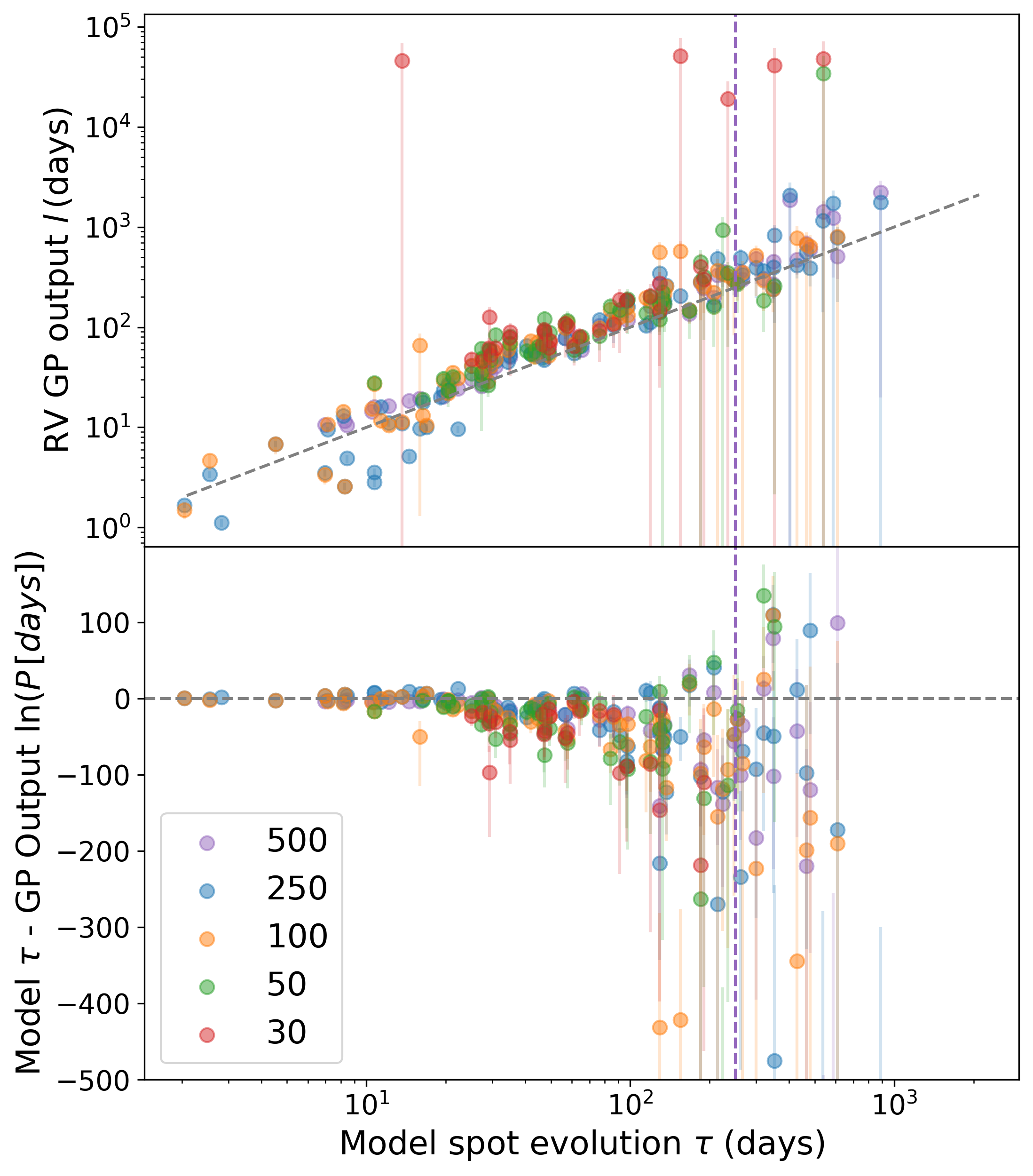}
    \caption{{This figure compares the solutions QP GP evolution time scale $l$ for different re-samplings of the RV data. The grey dashed line indicates the unity relation, and the grey vertical dashed lines indicate the total time span of the date of 250 days.}}
    \label{fig:RV_resamp_l}
\end{figure}

In Figure \ref{fig:RV_resamp_l} we compare the input model spot evolution time to the QP GP solutions for the evolution time scale in the re-sampled data sets. As seen with the rotation period solutions, there is greater scatter at smaller values of the spot evolution time. Further, in the 30 and 50 point data sets there are no converged solutions for evolutionary timescales below 8 and 10 days, respectively. While the full sample overestimates the evaluation timescale for mid-range values of spot evolution time, this overestimation improves at the lower sampling rate of 250 and 100 days, but then underestimates the evolution time in the 50 and 30 data point sets.

These differences between the time samples can be understood from the fact that the GP model is merely an approximation to the photometric and RV curves, a fact that becomes more noticeable with the greater sampling of the data. 

The spot model used to generate the data leads to some modest but noticeable high-frequency features (sharp changes in the data due to the emergence or disappearance of individual spots) that are apparent in the LC and RV curves, and that the QP GP struggles to reproduce. At the highest (500) time-sampling this actually leads to non-convergence in the cases of the shortest spot evolution times. 

At N=250 and 100, the MCMC converges, but the derived $l$ values are significantly lower than the injected spot lifetimes $\tau$. This is because the GP needs as much freedom as possible to reproduce these high frequency features, and it achieves this by having a shorter $l$. 

As the time-sampling decrease further (N=50 and 30), this no longer happens, as these high frequency features are no longer resolved, but we also lose the ability to recover short  evolution time-scales due to them being under-sample. These cases don't converge and aren't shown on the plot. For intermediate values of $\tau$, we observe that the values of $l$ are systematically larger than tau for low N. This is because with limited time sampling we can only place an upper limit on the rate at which the signal evolves. Once tau is long compared to the sampling it then becomes measurable by the QP GP. Note that all these considerations would apply to light curve data as well, but our light curves were not simulated with reduced time-sampling because typically light curve with sufficient precision are measured from space with the high cadence sampling that allows.

\begin{figure}
	\includegraphics[width=\columnwidth,trim={0 0cm 0 0cm},clip]{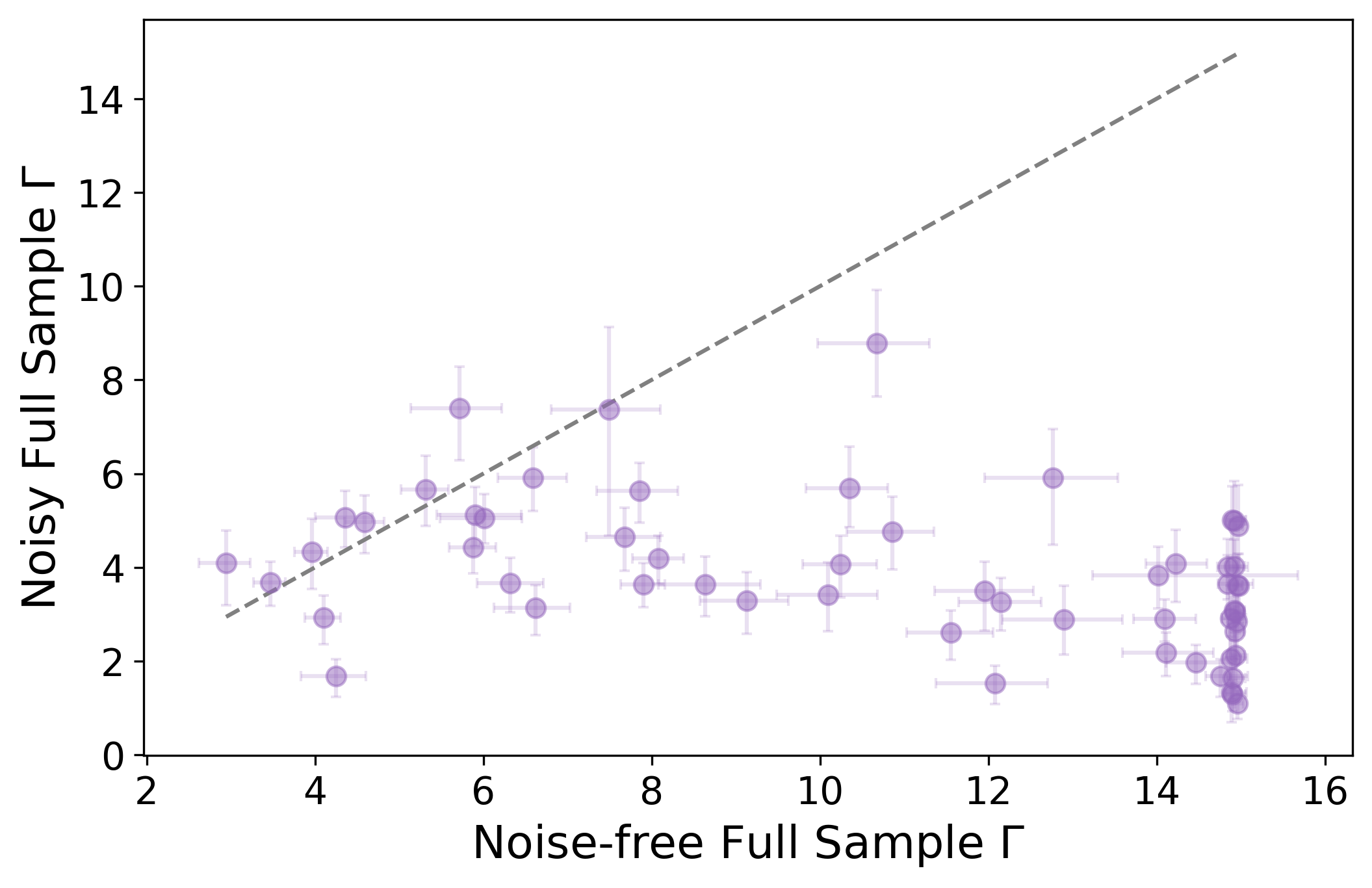}
    \caption{{This figure compares the solutions QP GP harmonic complexity $\Gamma$ for 'perfectly' sampled RV data with and without noise. The grey dashed line indicates the unity relation.} }
    \label{fig:RV_resamp_gamma_NF}
\end{figure}

{Figure \ref{fig:RV_resamp_gamma_NF} compares the $\Gamma$ values obtained for the noisy vs noise-free RV time-series, with the full 500-point sampling. The cluster of $Gamma$ values at the upper end of the prior range ($\Gamma=15$), which was present in the noise-free case, disappears in the noisy case. We speculate that, in the noise free cases, the GP model attempts to reproduce the small, sharp changes caused by the rapid emergence of individual spots, by adopting a short intra-periodic length scale (i.e.\ a large value of $\Gamma$). As soon as even small amounts of white noise are added in the data, these small changes are no longer discernible. There is no apparent correlation between the individual $\Gamma$ values obtained in the noisy and noise-free cases. This implies that, more than any physical property of the active regions, the noise properties of the data are the key factor driving the measured $\Gamma$ values.}

\begin{figure}
	\includegraphics[width=\columnwidth,trim={0 0cm 0 0cm},clip]{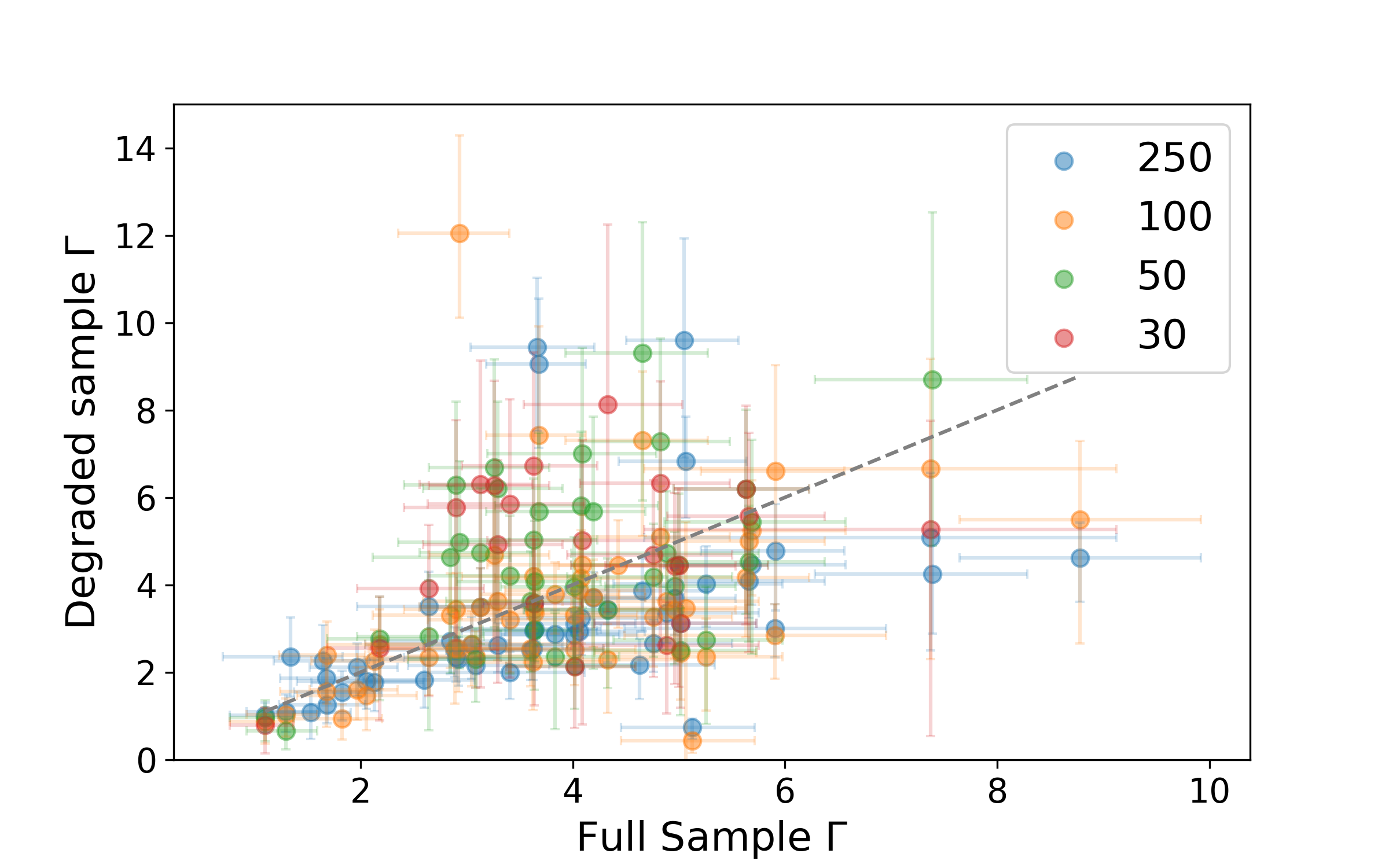}
    \caption{{This figure compares the solutions QP GP harmonic complexity $\Gamma$ for 'perfectly' sampled noisy RV data that with degraded time sampling. The grey dashed line indicates the unity relation.} }
    \label{fig:RV_resamp_gamma_NvN}
    
\end{figure}

{To assess the impact of sampling on $\Gamma$, Figure \ref{fig:RV_resamp_gamma_NvN} compares solutions to the noise-included full data set to the degraded data. Solutions below $\Gamma \sim 2.5$ across all sampling levels roughly agree whit the full sample solutions. Beyond this value, the scatter in $\Gamma$ values increases dramatically, regardless of the sampling. This again confirms that the $\Gamma$ parameter is predominately driven by noise, which is on the same scale across all samplings of data.  }

\section{Discussion and Conclusions}
\label{sec:DiscConc}
This work seeks to find what physical meaning, if any, can be attributed to the QP GP hyperparameters derived from photometric time series. In the context of the use of photometric data to inform stellar activity jitter models in RV data, we investigate the similarities and differences between the QP GP hyperparameters of photometric and RV data. Finally, we look at the impact of time sampling on RV data on the solution to a QP GP, and the impact this has on the inferred stellar properties. 

\subsection{The relation between QP GP hyperparameters and stellar properties}
To investigate the physical meaning of QP GP hyperparameters, we solve for them for a set of model light curves with varying rotation periods and spot evolution times with MCMC, and compare the output hyperparameters to the input stellar properties. The results from this showed almost perfect recovery of the input stellar rotation period, with increasing scatter for shorter periods due to the finite sampling of the data. This result is expected, and has been indicated previously. \cite{Angus2018} conducted a similar study to the work presented here focusing on the recovery of rotation periods from Kepler light curves by a QP GP, using a similar spot model to validate their results. The results in this work form a much tighter correlations with period, and fewer cases of the QP GP finding an alias of the true stellar rotation period due to the lack of noise included in our stellar model light curves. 

We also find a correlation between the spot evolution time and the QP GP length scale, consistent with behaviour demonstrated by \cite{Santos2021} between spot evolution time and light curve auto correlation decay timescales.  Toward longer spot evolution times the GP overestimates the evolution length scale, with increasing deviation and scatter when the spot evolution timescales approach the time span of the data. This correlation in tighter when the spot emergence and decay times are equal. This behaviour is a natural consequence of the fact that the GP kernel has only one parameter to describe the evolution of the periodic signal over time. If the spot emergence times are considerably shorter than the spot decay times, the simulated light and RV curves contain both shorter-term changes due to spot emergence, and slower changes due to spot decay. The best-fit GP evolutionary timescale is then a compromise between those values, and less tightly correlated to either. The longer-period single is the one that is preferentially found, as it is better sampled by the data. This indicates that the QP GP length scale can be used as an indicator of spot decay time and evolution.

The harmonic complexity term, $\Gamma$ does not have an intuitive physical interpretation, but does display some correlation with physical stellar properties. The $\Gamma$ parameter is weakly correlated with total spot number, though not at all with the average spot number. We also find $\Gamma$ to correlate moderately with rotation period and spot evolution time, although the latter is shown to be a spurious correlation. The simplicity of our spot model does not allow for a deeper exploration of the possible influences on $\Gamma$. Results from P21 indicate some influence of spot distribution on this hyperparameter. Future studies involving more complex stellar models with varying spot emergence patterns and differential rotation may reveal more about the usefulness on the $\Gamma$ hyperparameter in probing the physical properties of stars from their photometric and RV time series. 

\subsection{QP GP regression on photometry versus RVs}
We perform the same analysis on time series of RV data points to investigate if the same QP GP hyperparameters can be applied to both types of data. We show in this work that, although the period and the length scale hyperparameters match for both photometric and RV data, the harmonic complexity term does not: it is systematically higher in RV than photometric time series. This can be explained as the RV time series will behave as a combination of the photometric time series and its first derivative, creating a more complex change in the RV with time compared to the photometry. It is important to note, however, that the model curves used in this analysis can be considered in the `high complexity' regime for the QP GP harmonic complexity. \cite{PyanetiII} show that this behaviour of the harmonic complexity between a time series and its derivative is dependent on the complexity of the time series. A high harmonic complexity time series will have a derivative with a higher harmonic complexity. For time series data with low harmonic complexity, however, the harmonic complexity of the time series will be almost equally low. This explains the discrepancy between the findings here and those of \cite{Kosiarek2020}, who compare QP GP hyperparameters between Solar photometric and RV observations, and find a marginally lower harmonic complexity (their $\eta_4$ `length scale' is equivalent to $1/\Gamma$ ) in three of their RV data compared to the photometry, and the opposite in another, though in none of the sets are these values as drastically different as we observe in some of our models. We conclude from this that it is not appropriate to blindly use the harmonic complexity derived from photometry for analysis of RV time series. 

One common scenario not investigated in this work is the use of photometric data to constrain a GP on RVs taken months or years after the photometric observations, as happens in the case of RV follow-up of planet candidates from transit surveys.  This investigation is left to future work, as it requires a more sophisticated stellar model than the one used here. Such an investigation would require activity-cycle-like variations in addition to individual spot evolution and rotational modulation to best estimate the efficacy of using non-contemporaneous photometry.

\subsection{Comparative performance of the QP and QPC kernels}
The QP and QPC kernels behave almost identically for the recovery of stellar rotation period and spot evolution times in both photometric and RV data. In particular for photometric data, the QPC tended to have very small values of cosine fraction $f$, and so was largely behaving as a QP kernel. Since we allowed the harmonic complexity $\Gamma$ of the QPC kernel to vary in our analysis, it is possible that any variability present at half the rotation period was explained by higher $\Gamma$, rather than higher $f$. P21 explored this potential degeneracy by fixing the harmonic complexity and varying $f$ only. We did the same test, repeating our QPC analysis of the model light curves with $\Gamma$ set to $5.2$ (corresponding to the value of $\omega=0.31$ adopted by P21, where $\Gamma = 1/2\omega^2$). However, we found that this had almost no impact on the resulting distribution of $f$ values, which were almost identical to those obtained when $\Gamma$ was allowed to vary. In fact, the number of light curves with $f<0.1$ increased slightly. On the other hand, fixing $\Gamma$ significantly increased the scatter in the values of $l$ for $\tau>T/2$, and led to poorly constrained solution for the mean in around 10\% of our simulations.

In RV data, the fraction $f$ of the cosine component was higher on average, as was the harmonic complexity of the sine-squared term. Even so, the ability to recover periods and evolution timescales from RV data is largely the same with either kernel.

Both kernels gave rise to at least one significant outlier in our sample of 100 simulated stars, where the stellar period and evolution timescale $\tau$ were very different from the GP period and length scale $l$. The QPC kernel gave rise to even fewer of these outliers than the QP kernel, but the total numbers are very small in either case. 

We can also compare the QP and QPC kernels in terms of MCMC convergence. Both kernels led to non-converged solutions for some models: 2 and 10 in total for the QP kernel on light curves and radial velocities, respectively, compared with 2 and 1 for the QPC case. In real-world scenarios, though, the MCMC set up can be tuned to achieve convergence for each individual case. 

Overall, the QP and QPC kernels are comparable in performance, both having advantages and drawbacks. The QP kernel is simpler, with fewer parameters, and in most cases we do not find sufficient evidence of superior performance with the QPC kernel to justify the additional cost. The only advantage conferred by the QPC kernel on our sample appears to be improved MCMC convergence in around 10\% of the RV curves, though we were unable to identify any obvious reason why convergence was more difficult to reach in those particular cases. Future work with more realistic spot models and spot distribution patterns may shed more light on this. 

\subsection{The impact of {'realistic' noise and} time sampling {on QP GP analysis of RVs}}
To reflect the scenario where there is only RV data for the QP GP analysis, we degrade the time sampling of our RV data set to better reflect the observations from ground-based observatories. When it comes to the correlation of these degraded QP GP, the results show that the recovery in evolution time and period is limited to both the total time span of the data and the minimum separation of the data, the former limiting the ability to reproduce long evolution times, and the latter impacting the ability to sample the shorter period and evolution's times the larger this minimum separation becomes whit decreasing amounts of data. It is also apparent from the insets in Figure \ref{fig:RV_resamp_08} that the fewer data points give fewer constraints on the GP solution, making it increasingly more flexible. This means that there is more and more degeneracy between a GP-based activity model for the RVs and any potential planetary signal. This highlights the importance of adequate time sampling in RVs, both in total time span, and in sampling frequency.

\section*{Acknowledgements}

The authors would like to that Annelies Mortier, Helen Giles and the Oxford exoplanets group for their helpful suggestions and insights to this work. This project was carried out with support from STFC Consolidated Grant ST/S000488/1 (PI Balbus). SA also acknowledges support from the European Research Council (ERC) under under the European Union’s Horizon 2020 research and innovation programme (Grant agreement No. 865624).

\section*{Data availability statement}
The stellar model data can be found in the supplementary information to this paper, and the code used to generate them can be found here: \url{https://github.com/saigrain/pyspot} \citep{aigrain_suzanne_2021_5654179}. The code used for the QP GP analysis and time sampling is available on request. 





\bibliographystyle{mnras}
\bibliography{GPSimsBib} 



%
\appendix
\section{The non-converged solutions}
\label{sec:appA}
Across all of the studies in this work, there was always at least one model for which the MCMC did not converge to a single set of solutions for the QP {or QPC} GP hyperparameters. Further, the models that did not converge were not always the same. This section explores some of the non-convergence scenarios. 

\begin{figure*}
	\includegraphics[width=1\textwidth]{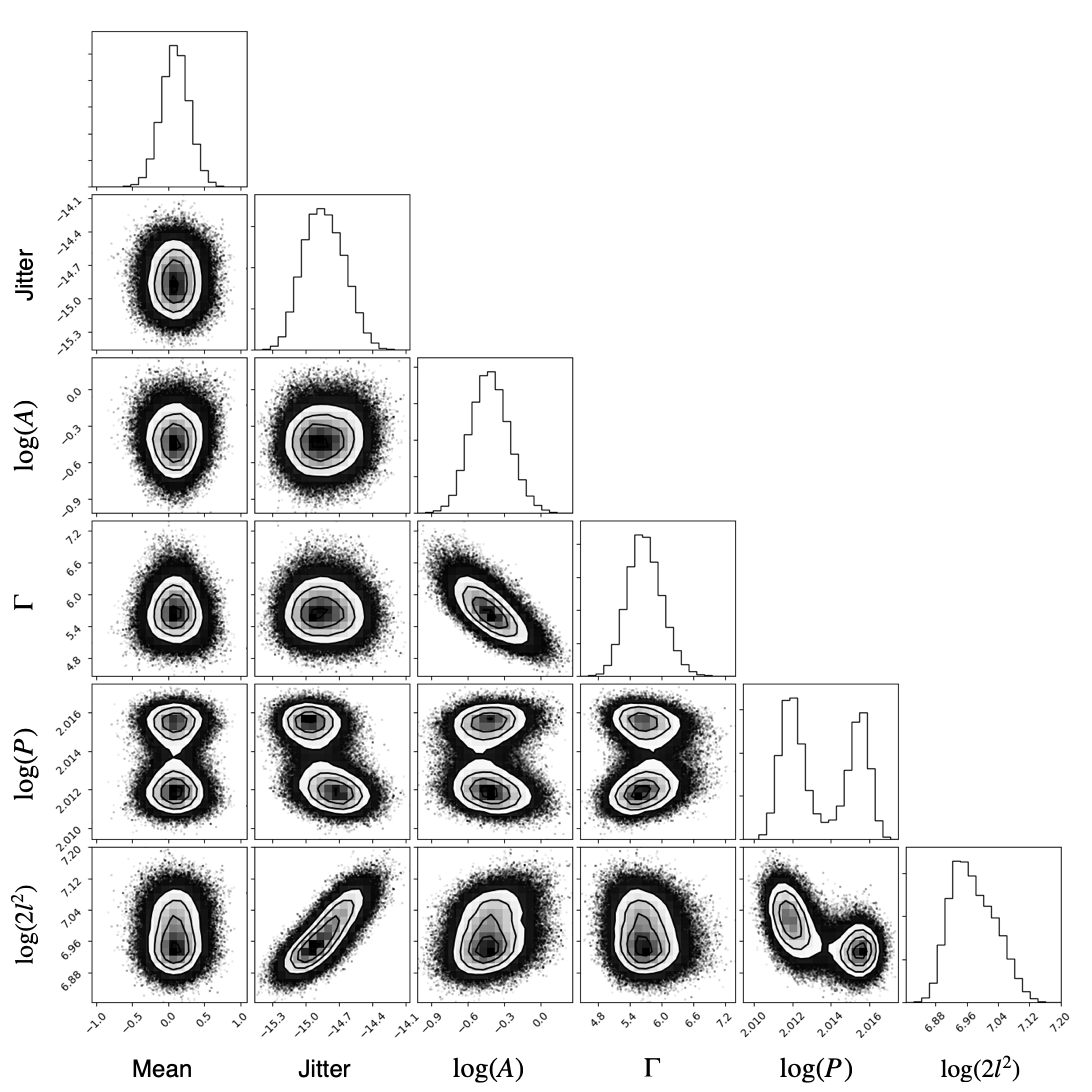}
    \caption{{An example of bimodal posterior distribution of a QP GP fit to a model light curve.}}
    \label{fig:Multimodal}
\end{figure*}

Most of the non-converged MCMC solutions gave multi-modal posterior distributions, an example of which is given in Figure \ref{fig:Multimodal}. This figure exemplifies what occurs in a majority of these multi-modal cases{all of the QPC non-converges cases, and most of the QP non-converged cases}: there are two equally probable solutions for the `jitter' white noise term and the harmonic complexity term. Given their equal likelihood, the MCMC is unable to prefer one solution over the other. Looking at the jitter vs $\Gamma$ panel in this figure we can see that they are correlated: the lower jitter values give solutions with higher harmonic complexity, and vice versa. Bimodality is seen in other hyperparameters as well, such as between l and period, or in period alone. The cases with bimodal posteriors have no similarities in their stellar parameters, and in the jitter-$\Gamma$ bimodal cases the solutions to the period and evolution timescale $l$ still match the input model period and spot evolution time, respectively, so can be considered reliable solutions despite not strictly meeting our convergence criteria due to the bimodality in the posteriors of the other parameters. Such a bimodality in posteriors is possible in real data too, and in such circumstances, so long as the likelihood are equal for both sets of parameters, they must be considered as equally likely given the data. In this work, however, given the lack of a single solution across all parameters these cases, for consistency across the results they are excluded from our analysis. 

\begin{figure*}
	\includegraphics[width=1\textwidth]{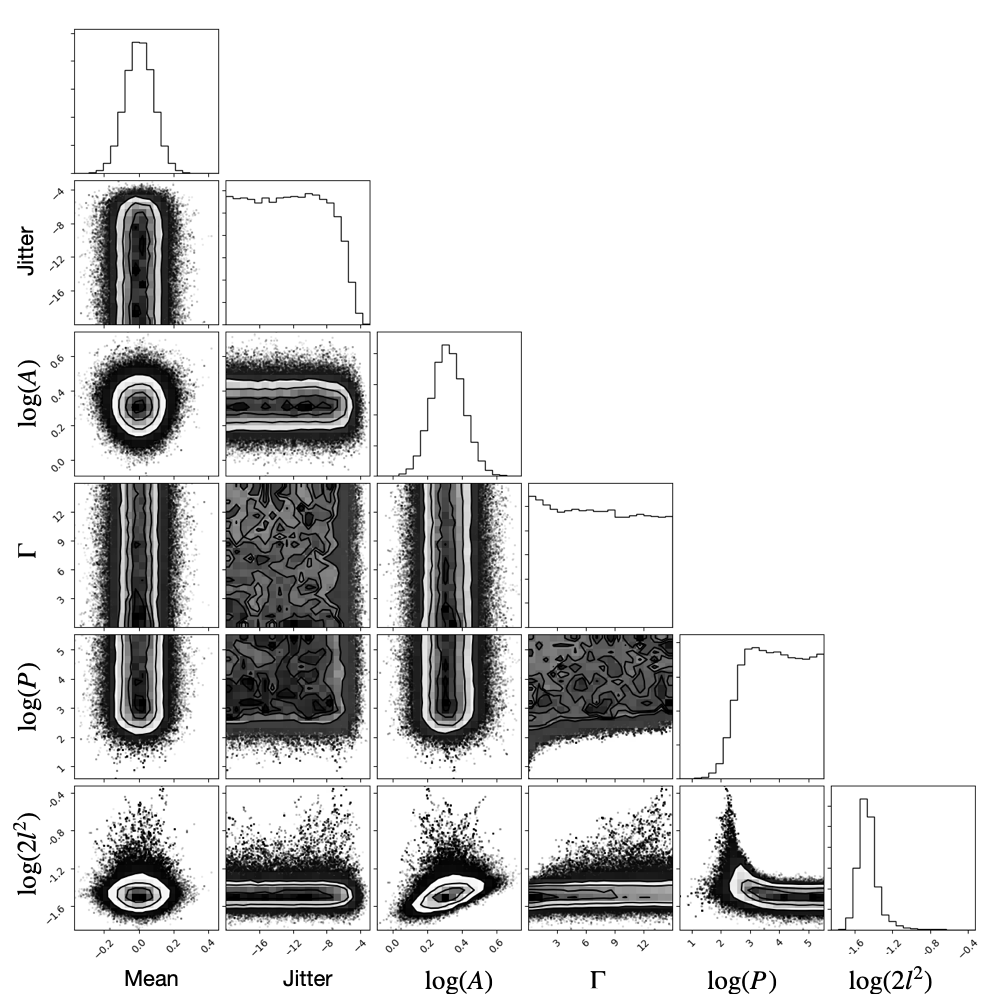}
    \caption{{An example of the posterior distribution of an unconstrained MCMC solution. } }
    \label{fig:unconstrained}
\end{figure*}

Some of the non-converged solutions for the QP analysis of the model RV data had unconstrained, poorly constrained posteriors or complex posterior distributions, which meant that the convergence criteria were not met in the 10000 step limit of our MCMC. As with the bimodal solutions, there is nothing particularly remarkable about the input stellar parameters of these models. The worst of the unconstrained solutions shown in figure \ref{fig:unconstrained}. While this model did have a relative short rotation period with a short evolution timescale, this was not an extreme case, and the QP GP was able to find converged solutions for even shorter rotations periods, and smaller ratios of rotation period to evolution timescale. Trying an MCMC with a different 'move' setup did not alter the convergence of these cases.

%


\bsp	
\label{lastpage}
\end{document}